\documentclass[fleqn,usenatbib]{mnras}

\usepackage{newtxtext,newtxmath}

\usepackage[T1]{fontenc}

\DeclareRobustCommand{\VAN}[3]{#2}
\let\VANthebibliography\thebibliography
\def\thebibliography{\DeclareRobustCommand{\VAN}[3]{##3}\VANthebibliography}

\usepackage{graphicx}	
\usepackage{amsmath}	

\title[Short title, max. 45 characters]{MNRAS \LaTeXe\ template -- title goes here}

\title[Influence of three parameters ...]{Influence of three parameters on maximum mass and stability of strange star under linear $f(Q)-$action}

\author[Santosh V. Lohakare et al.]{Santosh V. Lohakare, $^{1}$\thanks{E-mail: lohakaresv@gmail.com}
S. K. Maurya,$^{2}$\thanks{E-mail: sunil@unizwa.edu.om}
Ksh. Newton Singh,$^{3}$\thanks{E-mail: ntnphy@gmail.com}
B. Mishra$^{1}$\thanks{E-mail: bivu@hyderabad.bits-pilani.ac.in}
and Abdelghani Errehymy $^{4}$\thanks{E-mail: abdelghani.errehymy@gmail.com}
\\
$^{1}$ Birla Institute of Technology and Science-Pilani, Hyderabad Campus, Hyderabad 500078, India,\\
$^{2}$Department of Mathematics and Physical Science, College of Arts and Sciences, University of Nizwa, Sultanate of Oman\\
$^{3}$Department of Physics, National Defence Academy, Khadakwasla, Pune 411023, India\\
$^{4}$ Astrophysics Research Centre, School of Mathematics, Statistics and Computer Science, University of KwaZulu-Natal, Private Bag X54001, Durban 4000, South Africa
}

\date{Accepted XXX. Received YYY; in original form ZZZ}

\pubyear{2023}

\begin{document}
\label{firstpage}
\pagerange{\pageref{firstpage}--\pageref{lastpage}}
\maketitle

\begin{abstract}
    This study simulates strange stars in $f(Q)$ gravity with an additional source under an electric field using gravitational decoupling and the complete Gravitational Decoupling (CGD) technique. By employing the Tolman ansatz and the MIT bag model equation of state (EOS), we explore bounded star configurations derived from the $\theta_0^0 = \rho$ and $\theta_1^1 = p_r$ sectors within the CGD formalism. Our models are subjected to physical viability tests, and we analyze the impact of anisotropy and the electric charge parameter $E_0$ as well as the coupling parameters $\alpha$ and $\beta_1$. Comparisons are made with observational constraints, including GW190814, neutron stars PSR J1614-2230, PSR J1903+6620, Cen X-3 and LMC X-4. Notably, we achieve the presence of a lower "\textit{mass gap}" component by adjusting parameters $\alpha$ and $\beta_1$. Our models exhibit well-behaved mass profiles, internal regularity, and stability, with the absence of gravitational collapse verified through the Buchdahl--Andr\'{e}asson's limit. In addition, we present a detailed physical analysis based on three parameters, $\alpha$ (decoupling strength), $\beta_1$ ($f(Q)$--coupling) and $Q$ (surface charge). This study provides insights into the behavior of compact objects in $f(Q)$ gravity and expands our understanding of strange star configurations within this framework.
\end{abstract}



\section{Introduction}
    Several independent cosmological observations indicate that the Universe is accelerating \citep{Riess_1998_116, Perlmutter_1998_517, Suzuki_2012_746}. The cosmological constant ($\Lambda$) in the Einstein-Hilbert action integral allows the General Theory of Relativity (GR) to represent the recent accelerated expansion of the Universe. Einstein's GR has achieved considerable success, but one of its challenges has been the cosmological observations posed by dark energy and dark matter. Typically, two approaches address this bizarre issue: the first approach is to change the matter sector by introducing more dark energy components into the energy budget of the Universe, and the second approach is to extend the geometrical part of GR. The second approach involves the extension of Einstein-Hilbert action. The most common geometrical extension of GR is the $f(R)$ gravity, which is based on curvature. Another approach is to extend the geometrical part through torsion and nonmetricity. The torsion-based gravitational theories, which is equivalent to GR, is known as the Teleparallel Equivalent of General Relativity (TEGR) \citep{Einstein_1928_17, Hayashi_1979_19, Sauer_2006_33} whereas in the non-metricity based it is known as Symmetric Teleparallel Equivalent of General Relativity (STEGR) \citep{Nester_1999, Jimenez_2018_039, Jimenez_2018_98}. The TEGR gravity is based on torsion instead of the curvature of space-time. This theory is developed in flat space-time with torsion, where the primary variables are the tetrad and spin connection when using the tetrad language. The vanishing of the curvature and non-metricity tensor limits the spin connection. These constraints enable the selection of the Weitzenb\"ock connection, where all components of the spin connections disappear, leaving simply the tetrad components as the fundamentals. This choice is considered a gauge selection in TEGR. It does not impact physics since any other compatible choice with the constraint of teleparallelism will lead to the same action except for a surface term.

    In STEGR, nonmetricity defines gravity rather than curvature and torsion. Given the teleparallelism bounds, a coincidence gauge can be chosen in this theory, which establishes just the metric tensor as a fundamental variable. Further extension of STEGR  is the $f(Q)$ gravity, which is quite comparable to $f (R)$ gravity \citep{Jimenez_2018_98, Heisenberg_2019_796}. Several aspects of $f(Q)$ gravity are available in the literature (\citep{Jimenez_2020_101, Anagnostopoulos_2021_822, Barros_2020_30, Flathmann_2021_103}). The extended theories of gravity have given significance in finding the astrophysical and cosmological properties of the Universe. In the general class of $f(Q)$ gravity, the propagation velocity and potential polarization of gravitational waves across Minkowski space-time were investigated in (\citet{Hohmann_2019_99}). Consequently, gravitational wave polarization has a significant influence on the strong-field behavior of gravitational theory (\citet{Soudi_2019_100}). The $f(Q)$ theory has been studied in a wide range of contexts, including those related to late-time acceleration with observational data sets \citep{Lazkoz_2019_100} bouncing \citep{Bajardi_2020_135}, black holes \citep{Ambrosio_2022_105}, and the development of the growth index in matter perturbations \citep{Khyllep_2021_103}. Several $f(Q)$ parametrizations have been studied to incorporate observational data constraints (\citet{Ayuso_2021_103}). 

    Gravitational waves and black hole shadows have strengthened the study of compact objects such as pulsars, neutron stars (NS), and quark stars \citep{Abbott_2016_116, Akiyama_2019_875}. Einstein's classical GR has produced stellar models that match data in different contexts. Stars with densities equal to $10^{16} \text{g cm}^{-3}$ have continued to be explained by GR, which accounts for their mass-to-radius ratios, compactness, and redshifts. The electromagnetic field significantly influences the development and stability of compact objects. The most crucial components for lowering the gravitational force are electromagnetic forces. Compact star objects need a massive charge to maintain strength and overcome the tremendous gravitational pull. According to Bekenstein's analysis \citep{Bekenstein_1971_4}, charge acts as a Coulomb repulsive force and maintains the system's stable state. In Ref. \citet{Esculpi_2010_67}, it is shown that charge and anisotropy exhibit repulsive behavior in extending this study to anisotropic matter configuration. In Ref. \citet{Rahaman_2012_72}, an analysis of the effect of charge on neutron star structure in an anisotropic fluid configuration has been shown. To solve the Einstein-Maxwell equations, \citet{Maurya_2015_75} considered the baryonic matter with the charge and obtained new charged stellar models where the matter variables rely on the electromagnetic field. The dynamical evolution of compact stellar objects was studied in Ref. \citet{Sharif_2016_48}. They discovered that charge plays a significant role in the evolution of star instability. The existence of compact structures with anisotropic matter distributions in the framework of $f (R, T)$ theory in light of modified gravitational theories has been investigated in Ref. \citet{Maurya_2019_100}. 

    A neutron star obeying a stiff causal EOS within the context of $f (R)$ gravity can have a maximum mass that is determined by computational techniques and by the GW190814 event, according to \citet{Astashenok_2020_811}. On the basis of modified gravity, this analysis accounted for a neutron star's mass, $\approx$ 3 M$_{\odot}$, as the most severe upper constraint. However, Astashenok and his collaborators Odintsov and Capozziello produced the most influential work on compact star models (including neutron stars) in the $f(R)$-gravity theory. \citet{Astashenok_2020_498} used the GM1 EOS to analyze the rotating neutron star in $f(R)$-gravity with axions. They also studied a realistic supermassive neutron star in $f(R)$-gravity, comparing their findings with those obtained with the static star \citep{Astashenok_2020_493}. This EOS has causal limits within the mass-gap area, with masses $M > 2.5$ M$_{\odot}$ but below the 3 Solar Masses. In this context, Capozziello and his collaborators \citep{Astashenok_2021_816, Astashenok_2020_811, Nashed_2021_81} discussed the neutron star model of $f(R)$ gravity theory and its anisotropic star and mass-radius relation. In perturbative $f(R)$--gravity theory, the stable neutron star was generated using FPS and the SLy EOS with logarithmic and cubic corrections \citep{Astashenok_2013_2013_040}. The SLy EOS assessed the star's minimum radius as 9 km and its maximum mass as 1.9 M$_{\odot}$. See Ref. \citep{Astashenok_2015_2015_001}, which offers extreme neutron stars in the extended theory of gravity (in particular, the $f(\mathcal{G})$ and $f(R)$ theories), providing that the large realistic star with the maximum mass $M > 4$ M$_{\odot}$ and the radius 12 - 15 km may be obtained. Under several EOS by taking $f(R) = R + \alpha R^2$, the realistic relativistic star in $f(R)$ theory is examined \citep{Astashenok_2017_34}.

    In recent investigations, the concept of GD has been incorporated into several modified frameworks. As in the standard 4D classical gravity theory, GD allows for anisotropisation of seed solutions, thus providing a mechanism to study the impact of anisotropic stresses in compact objects. The minimal and extended geometric deformation methods \cite{Ovalle_2017_95,Ovalle_2019_788} were utilized to model a compact star in different aspects \cite{MGD1, MGD7, MGD8, MGD9, MGD13, MGD27, MGD2, MGD10, MGD26, MGD14, MGD15, MGD16, MGD17, MGD20, MGD21, MGD22, MGD30, MGD31}. This work showed that synergistic contributions from the decoupling parameters lead to higher neutron star masses. Compact star objects with masses aligned with the GW190814 event in $f(Q)$ gravity have been identified in \citet{Maurya_2022_2022_003}. The metricity parameter and the deformation constant viably and steadily determine models of compact stellar systems. The minimal gravitational decoupling and extended gravitational decoupling, stellar masses above 2 M$_{\odot}$, have been successfully accounted for without requiring exotic matter distributions \citet{Maurya_2022_2022_003, Maurya_2022_70, Maurya_2022_925}. Modeling compact objects utilizing gravitational decoupling is successful in classical and modified gravity theories, including modeling black holes and lensing \citep{Ovalle_2013_88}. The extended gravitational decoupling method combined with a complete geometric deformation (CGD) method is used to model strange stars within the symmetric teleparallel formalism with the electromagnetic field. This study examines the nonmetricity and the effect of coupling constant on the maximum allowable masses and radii for self-gravitating compact stars PSR J1614-2230, PSR J1903+6620, Cen X-3 and LMC X-4. We compare our findings with the observational constraints found in GW190814.

    The article is organized as follows: We present in Section \ref{sec2} the equations for $f(Q)$ gravity with an additional source under an electric field and obtain the modified Tolman-Oppenheimer-Volko (TOV) equations for spherically symmetric solutions. We impose boundary conditions in Section \ref{sec3} to match the interior with the exterior solution. We discuss the gravitationally decoupled solution of $f(Q)$ gravity in Section \ref{sec4}. The model is consistent with a realistic compact star based on the physical analysis and astrophysical implications discussed in Section \ref{sec5}. Subsequently, in Section \ref{sec6}, the $M-R$ curves are used to analyze the mass-radius constraints. The importance of energy exchange in extended gravitational decoupling is addressed in Section \ref{sec7}. Using small radial perturbations and Buchdahl limits of gravitational collapse, Section \ref{sec8} examines the stability of each configuration. Section \ref{sec9} summarizes the findings and provides an outlook.

\section{Field equations for \texorpdfstring{$\lowercase{f}(Q)$}{} gravity with extra source under electric field}\label{sec2}

    A brief description of STEGR will be discussed here. A manifold ($\mathcal{M},\, g_{ij},\, \Gamma^{k}_{ij}$) satisfying the generic metric-affine theory, where $g_{ij}$ is a metric tensor with signature ($-1,\, +1,\, +1,\, +1$), and $\Gamma^{k}_{\,\,\,ij}$ denotes any arbitrary affine connection. Torsion tensor is defined as
        
    \begin{eqnarray} 
            && \hspace{-5.3cm} T^{k}_{\,\,\,ij}=2\Gamma^k_{\,\,\,[ij]}=\Gamma^{k}_{\,\,\,ij}-\Gamma^{k}_{\,\,\,ji} \,,  \label{eq.1}
    \end{eqnarray}
and curvature,
    \begin{eqnarray} \label{eq.2}
        && \hspace{-3.7cm} R^{l}_{kij}=\partial_i \Gamma^{l}_{jk}-\partial_j \Gamma^{l}_{ik} + \Gamma^{m}_{jk} \Gamma^{l}_{im} - \Gamma^{m}_{ik} \Gamma^{l}_{jm}
    \end{eqnarray}
the nonmetricity of the connection is defined by
    \begin{eqnarray}\label{eq.3}
        && \hspace{-3.2cm}Q_{k i j}=\bigtriangledown_{_k}\,g_{i j}=\partial_i\, g_{jk}-\Gamma^l_{\,\,\,i j}\, g_{l k}-\Gamma^l_{\,\,\,i k} \,g_{j l}\,,
    \end{eqnarray}
where $\nabla_{k}$ is the covariant derivative, so the affine connection can be shown as
    \begin{eqnarray}  \label{eq.4}
        && \hspace{-6.9cm}\Gamma^k_{i j}=\mathring{\Gamma}^{k}_{\,\,\,ij} - S^{k}_{\,\,\,ij}\,,
    \end{eqnarray} 
where $\mathring{\Gamma}^{k}_{\,\,\,ij}$ is Levi-Civita connection, is defined as
    \begin{eqnarray} \label{eq.5}
        &&\hspace{-4.2cm} \mathring{\Gamma}^{k}_{\,\,\,ij} = \frac{1}{2}g^{ kl}\left(\partial_i \,g_{l j}+\partial_ j \,g_{l i}-\partial_l \,g_{i j}\right)\,,
    \end{eqnarray}
and the disformation $L^{k}_{ij}$ is defined by
    \begin{eqnarray} \label{eq.6}
        &&\hspace{-4cm} L^{k}_{ij} \equiv \frac{1}{2} g^{kl} \left(-Q_{ilj}-Q_{jli}-Q_{lij} \right)=L^{k}_{ji}
    \end{eqnarray}
As it relates to the nonmetricity tensor, the superpotential is defined as:
    \begin{eqnarray} \label{eq.7}
        && \hspace{-1.5cm}P^{kij}=-\frac{1}{4} Q^{kij} + \frac{1}{2} Q^{(ij)k}+\frac{1}{4}(Q^k -\tilde{Q}^k) g^{ij}-\frac{1}{4}\delta^k {{}^({}^i}Q^{j)},
    \end{eqnarray}
where the vectors $Q^k$ and $\tilde{Q}^k$ are two independent traces from the nonmetricity tensor $Q_{kij}$ as
    \begin{eqnarray} \label{eq.8}
        && \hspace{-5.3cm}Q_{k}\equiv Q_{k\,\,\,~i}^{\,\,~i},~~~~~~~~\; \tilde{Q}^k=Q^{\,\,\,\,ki}_{i}.
    \end{eqnarray}
Finally, the definition of the nonmetricity scalar is as follows:
    \begin{eqnarray} \label{eq.9}
        && \hspace{-3.7cm}Q=-g^{ij}\left(L^{k}_{lj} L^{l}_{ik}-L^{l}_{il} L^{k}_{ij}\right)=-Q_{kij} P^{kij}
    \end{eqnarray}

The affine connection has the following form because it has no torsion and no curvature; it can be more explicitly parameterized by a set of functions, such as
    \begin{eqnarray} \label{eq.15}
        && \hspace{-6.4cm}\Gamma^k_{\,\,\,i j}=\left(\frac{\partial x^k}{\partial\xi^l}\right)\partial_i \partial_j \xi^l.
    \end{eqnarray}
where $\xi^l$ is arbitrary space-time position functions are considered, using a general coordinate transformation, We always have the option of selecting a coordinate of the form $\xi^l=\xi^l (x^i)$. The general affine connection has a $\Gamma_{ij}^{k} = 0$. According to \citet{Jimenez_2018_98}, this coordinate is called the coincident gauge. As a result, all covariant derivatives in the coincident gauge correspond to ordinary derivatives in the standard gauge, and the nonmetricity equation (\ref{eq.3}) is simplified to
    \begin{eqnarray} \label{eq.16}
        && \hspace{-7.4cm}Q_{k i j}=\partial_k \,g_{i j},
    \end{eqnarray}

As a result, the metric simplifies the computation. Diffeomorphism invariance is no longer present for this action, except for the standard General Relativity. It is possible to avoid a problem using a covariant formulation of $f(Q)$ gravity. It can use the covariant formulation by determining the affine connection in the absence of gravity before choosing the affine connection in equation (\ref{eq.15}).

\subsection{\texorpdfstring{$f(Q)$}{} Gravity}
Gravitational action integral in symmetric teleparallel $f (Q)$ gravity \citep{Jimenez_2018_98, Zhao_2022_82}
as follows by adding a second Lagrangian $\mathcal{L}_\theta$ and electromagnetic field
    \begin{eqnarray} 
        && \hspace{-3cm}\mathcal{S}=\int{\left(\frac{1}{2}\,f(Q)+\mathcal{L}_m + \mathcal{L}_e + \alpha \,\mathcal{L}_\theta\right)}\,\sqrt{-g}~d^4x \nonumber\\&& \hspace{-2.5cm} +\int{\left(\lambda^{\, kij}_l R^{l}_{\, kij} + \tau^{\,\,\, ij}_k T_{\,\,\, ij}^{k}\right)} ~d^4x\, , \label{eq.10}
    \end{eqnarray}
where $g$ is the determinant of the metric tensor i.e., $g={\begin{vmatrix}g_{ij}\end{vmatrix}}$, $\mathcal{L}_m$ is the description of the matter Lagrangian density, $\mathcal{L}_e$ denotes the matter Lagrangian for electromagnetic field and $\alpha$ is a decoupling constant. The Riemann and torsion tensors can be represented by $R^{l}_{\, kij}$ and $T_{\,\,\, ij}^{k}$, respectively. $\lambda^{\, kij}$ and $\tau^{\,\,\, ij}_k$ are two Lagrange multipliers that ensure that the connection $\Gamma_{\,\,\, ij}^k$ is flat and symmetric, meaning that the values of $R^{l}_{\, kij}$ and $T_{\,\,\, ij}^{k}$ are both zero.

The gravitational field equations are constructed by modifying the modified Einstein-Hilbert action (\ref{eq.10}) about the metric tensor $g_{i j}$,
    \begin{eqnarray} \label{eq.11}
        && \hspace{-0.75cm}\frac{2}{\sqrt{-g}}\bigtriangledown_k\left(\sqrt{-g}\,f_Q\,P^k_{\,\,\,\,i j}\right)+\frac{1}{2}g_{i j}f 
        +f_Q\big(P_{i\,k l}\,Q_j^{\,\,\,\,k l}-2\,Q_{k l i}\,P^{k l}_{\,\,\,\,\,j}\big) \nonumber\\
        && \hspace{0.3cm}=- \left(T^{\text{eff}}_{i j}+E_{i j}\right), 
        ~~~\text{where}~~T^{\text{eff}}_{i j}=\big(T_{i j}+\alpha \,\theta_{i j}\big),~~~~
    \end{eqnarray}
where $f_Q=\frac{\partial f}{\partial Q}$, and $T_{i j}$ is the energy-momentum tensor, electromagnetic field tensor $E_{ij}$ and extra source $\theta_{i j}$, which general form could be written as:
    \begin{eqnarray} \label{eq.12}
        && \hspace{-1.8cm}T_{ij}=-\frac{2}{\sqrt{-g}}\frac{\delta\left(\sqrt{-g}\,\mathcal{L}_m\right)}{\delta g^{i j}},\,~~E_{ij}=\frac{2}{\sqrt{-g}}\frac{\delta\left(\sqrt{-g}\,\mathcal{L}_e\right)}{\delta g^{i j}},\nonumber\\
        && \hspace{0.4cm}~~~~~~~~\text{and}~~~~~\theta_{i j}=-\frac{2}{\sqrt{-g}}\frac{\delta\left(\sqrt{-g}\,\mathcal{L}_\theta\right)}{\delta g^{i j}},~~~~
    \end{eqnarray}
Again, by varying the action (\ref{eq.10}) with the connection, we may have, as a result,
    \begin{eqnarray} \label{eq.13}
        && \hspace{-4.8cm}\bigtriangledown_i \bigtriangledown_j \left(\sqrt{-g}\,f_Q\,P^k_{\,\,\,\,i j}+H^{k}_{\,\,\,\,i j}\right)=0,
    \end{eqnarray}
where $H^k_{\,\,\,\,i j}=-\frac{1}{2} \frac{\delta \mathcal{L}_m}{\delta \Gamma^k_{\,\,\,i j}}$, denotes the hyper momentum tensor density. We may also extract the additional constraint over the connection, $\bigtriangledown_i \bigtriangledown_j (H^k_{\,\,\,\,i j})=0$ according to equation (\ref{eq.13}),
    \begin{eqnarray} \label{eq.14}
        && \hspace{-5.8cm}\bigtriangledown_i \bigtriangledown_j \left(\sqrt{-g}\,f_Q\,P^k_{\,\,\,\,i j}\right)=0.
    \end{eqnarray}

We are searching for gravitationally decoupled solutions for compact objects with $f(Q)$ gravity. In terms of theories of gravity, static spherically symmetric space-time is a fundamental assumption that provides us with an understanding of many aspects of astronomy. In static spherical symmetry space-time, spherically symmetric coordinate systems make sense. Here we have taken the spherically symmetric metric, 
    \begin{eqnarray} \label{eq.17}
        && \hspace{-2.5cm}ds^2=-e^{\Phi(r)}dt^2+e^{\lambda(r)}dr^2+r^2 (d\theta^2+\sin^2{\theta} \,d\phi^2), 
    \end{eqnarray}
In this study, we will focus on the anisotropic matter distribution, in which the effective energy-momentum tensor can be described as follows:
    \begin{eqnarray} \label{eq.18}
        &&\hspace{-2.2cm}T^{\text{eff}}_{ij} = \left(\rho^{\text{eff}} + p^{\text{eff}}_t\right)u_{i} \,u_{j} + p^{\text{eff}}_t\,g_{ij}+ \left(p^{\text{eff}}_r -p^{\text{eff}}_t\right)v_{i}\,v_{j},
    \end{eqnarray}
The fluid's four-velocity vector is $u_{i}$, and the effective density is $\rho^{\text{eff}}$. In addition to $v_{i}$, which represents the radial unit space-like vector, $p^{\text{eff}}_r$ and $p^{\text{eff}}_t$ represent the effective radial pressure and tangential pressure, respectively, in the direction of $u_{i}$ and $v_{i}$. However, the variables $u_i$ and $v_i$ fulfil the relationships $u^iu_i=-1$,  $v^iv_i=1$, and $u^iv_i=0$. On the other hand, the electromagnetic stress tensor can be described as,
    \begin{eqnarray} \label{eq.19}
        &&\hspace{-4.2cm}E_{ij}=\frac{2}{\kappa^2} \left(-F^{n}_{i}F_{j\,n} + \frac{1}{4}{g_{ij}} F_{k\,n}F^{k\,n} \right), 
    \end{eqnarray}
where $\kappa^2=\frac{8\pi G}{c^4}$, we considering $c=8\pi G=1$, here $G$ is Newtonian gravitational constant. The anti-symmetric electromagnetic field tensor $F_{i j}$ given in equation (\ref{eq.19}) is characterized as 
    \begin{eqnarray} \label{eq.20}
        &&\hspace{-6.5cm}F_{ij}=\nabla_{i}\,A_{j}-\nabla_{j}\,A_{i}
    \end{eqnarray}
for which Maxwell's equations have been satisfied,
    \begin{eqnarray} \label{eq.21}
        &&\hspace{-3cm}F_{i j,k} + F_{j k,i}+F_{k i, j}=0,~~\text{with}~~~
        F^{ik}\,_{;\,k} = \frac{J^i}{2}
    \end{eqnarray}
where $J^i$ is the electromagnetic 4-current vector. This can be expressed as
    \begin{eqnarray} \label{eq.22}
        &&\hspace{-6.2cm}J^{i}=\frac{\sigma}{\sqrt{g_{00}}}\,\frac{dx^i}{dx^0}=\sigma\,u^i, 
    \end{eqnarray}
The charge density, denoted by $\sigma=e^{\lambda/2}\, J^{0}(r)$, is measured in radians and is fundamental to understanding the electromagnetic 4-current $J^i$. In the case of a static matter distribution with spherical symmetry, only one component of the electromagnetic 4-current $J^i$ is non-zero and is a function of the radial distance $r$. Furthermore, only the $F^{01}$ and $F^{10}$ components of the electromagnetic field tensor, which describe the radial constituent of the electric field, are non-zero as defined by (\ref{eq.19}). These two components are related by the formula $F^{01}=-F^{10}$. This relation is an example of the mathematical expression of the conservation of electric charge in a system. As a result of this relation, the electric field has no tangential component, and the field is always directed radially outward or inward, depending on the charge sign. According to equations (\ref{eq.21}) and (\ref{eq.22}), the electric field is composed of the following constituents:
    \begin{eqnarray} \label{eq.23}
        &&\hspace{-5.5cm}F^{01} = - F^{10}=\frac{q}{r^2}\,e^{-(\Phi+\lambda)/2}
    \end{eqnarray}
According to the relativistic Gauss law and corresponding electric field $E$, this electric charge can be characterized directly by the quantity $q(r)$ that represents a spherical system of radial coordinates $r$ as follows:
    \begin{eqnarray} 
        &&\hspace{-3.5cm}q(r)=\frac{1}{2}\int^{r}_{0}{\sigma\,r^2\,e^{\lambda/2} dr}=r^2\,\sqrt{-F_{10}\,F^{10}}\label{eq.24} \\
        &&\hspace{-3.5cm}E^2=-F_{10}\,F^{10}=\frac{q^2}{r^4}. \label{eq.25}
    \end{eqnarray}
For the metric equation (\ref{eq.17}), we can calculate the non-metricity scalar as follows:
    \begin{eqnarray} \label{eq.26}
        &&\hspace{-5.6cm}Q=-\frac{2 e^{-\lambda(r)} \left(1+r \Phi^\prime(r)\right)}{r^2}, 
    \end{eqnarray}
In this expression of $Q$, here, $'$ represents the derivative over $r$ only, and $Q$ is based on zero affine connections. Based on the equations of motion (\ref{eq.11}) for the anisotropic fluid (\ref{eq.18}), the independent components are as follows:
    \begin{eqnarray}
        && \hspace{-0.65cm}\rho^{\text{eff}} +E^2 =\frac{f(Q)}{2}-f_{Q}\Big[Q+\frac{1}{r^2}+\frac{e^{-\lambda}}{r}(\Phi^\prime+\lambda^\prime)\Big],\label{eq.27}\\
        && \hspace{-0.65cm} p^{\text{eff}}_r-E^2=-\frac{f(Q)}{2}+f_{Q}\Big[Q+\frac{1}{r^2}\Big],\label{eq.28}\\
        && \hspace{-0.65cm} p^{\text{eff}}_t+E^2=-\frac{f(Q)}{2}+f_{Q}\Big[\frac{Q}{2}-e^{-\lambda} \Big\{\frac{\Phi^{\prime \prime}}{2}+\Big(\frac{\Phi^\prime}{4}+\frac{1}{2r}\Big) (\Phi^\prime-\lambda^\prime)\Big\}\Big],\label{19}\nonumber \label{eq.29}\\ 
    \end{eqnarray}
Suppose we assume that the affine connection in this coordinate system has a value of zero. We further require that the $f(Q)$ theory has vacuum solutions, which is to say that $\zeta_{r\theta} = 0$, then one can read out the off-diagonal component of equation (16), which states that the solutions to this equation are.
    \begin{eqnarray} \label{eq.30}
        && \hspace{-5cm} \zeta_{r\theta}=\zeta_{\theta r}=\frac{\cot \theta}{2}\,Q^\prime\,f_{QQ}=0. \label{20}
    \end{eqnarray}
Combining the solutions of these equations with the diagonal elements of the equation (general field eq.), we obtain $f_{QQ} = 0$. According to the findings above, EOMs will be inconsistent if $f(Q) = Q^2$ is selected at the outset. Metric equation (SSS metric) with affine connection $\Gamma_{ij}^k=0$ is thus not a solution of the equation of motions for those theories in which $f(Q)$ are not linear functions of $Q$. This does not imply that there are no static spherically symmetric vacuum solutions in $f(Q)$ theory; instead, it indicates that the spherically symmetric coordinate system is incompatible with the coincident gauge. In order to determine the functional form of $f(Q)$,
    \begin{eqnarray} \label{eq.31}
        &&\hspace{-2.5cm}f_{QQ}=0~\Longrightarrow~f_{Q}=-\beta_1~\Longrightarrow~f(Q)=-\beta_1\,Q-\beta_2 ,
    \end{eqnarray}
where $\beta_1$ and $\beta_2$ are constants. When equations (\ref{eq.26}) and (\ref{eq.31}) are inserted into equations (\ref{eq.27})--(\ref{eq.29}), the following exact formulations of equations of motion are obtained
    \begin{small}
    \begin{eqnarray}
        && \hspace{-0.8cm}\rho^{\text{eff}}+E^2 = \frac{1}{2\,r^2} \Big[2\, \beta_1+2\, e^{-\lambda}\, \beta_1  \left(r\, \lambda^\prime-1\right)-r^2 \,\beta_2 \Big], \label{eq.32}\\
        && \hspace{-0.8cm}p^{\text{eff}}_r-E^2=\frac{1}{2\,r^2} \Big[-2\, \beta_1+2\, e^{-\lambda} \,\beta_1  \left(r\, \Phi^\prime+1\right)+r^2\, \beta_2\Big], \label{eq.33}\\
        && \hspace{-0.8cm}p^{\text{eff}}_t+E^2=\frac{e^{-\lambda}}{4\,r} \Big[2\, e^{\lambda}\, r\, \beta_2 +\beta_1\,  \left(2+r \Phi^\prime\,\right) \left(\Phi^\prime-\lambda^\prime\right) +2\, r\, \beta_1 \, \Phi^{\prime \prime}\Big],~~~~\label{eq.34}
    \end{eqnarray}
    \end{small}
The vanishing of the covariant derivative of the effective energy-momentum tensor is $\bigtriangledown^i T^{\text{eff}}_{ij}=0$, provides,
    \begin{eqnarray}
        && \hspace{-1.9cm}-\frac{\Phi^\prime}{2}(\rho^{\text{eff}}+p^{\text{eff}}_r)-(p^{\text{eff}}_r)^{\prime}+\frac{2}{r}( p^{\text{eff}}_{t}-p^{\text{eff}}_r)+\frac{2qq'}{r^4}=0.~~~\label{eq.35}
    \end{eqnarray}
In $f(Q)$ gravity, the equation (\ref{eq.35}) is known as a TOV equation, with $f(Q)=-\beta_1 Q-\beta_2$. Following that, we want to use gravitational decoupling using the CGD approach to solve the system of equations (\ref{eq.32})-(\ref{eq.34}) for the compact star model. To put this into action, we modify the gravitational potentials $\Phi(r)$ and $\lambda(r)$ by inserting two arbitrary deformation functions through the decoupling constant $\alpha$ as follows:
    \begin{eqnarray}
        && \hspace{-5.4cm}\Phi(r) \longrightarrow H(r)+\alpha\, \xi(r)~ \label{eq.36}\\
        && \hspace{-5.4cm}e^{-\lambda(r)} \longrightarrow W(r)+\alpha\, \Psi(r).~~ ~~ \label{eq.37}
    \end{eqnarray}
where $\xi(r)$ and $\psi(r)$ are the geometric deformation functions for the temporal and radial metric components, respectively. When $\alpha = 0$, the classic $f(Q)$ gravity theory is usually recoverable. Because we are using the CGD method to solve the field equations, both deformation functions must be non-zero, i.e., fix $\xi(r) \ne 0$ and $\psi(r) \ne 0$. That means the metric function's radial and temporal components are impacted. The decoupled system (\ref{eq.32})-(\ref{eq.34}) is divided into two subsystems by the transformations (\ref{eq.36}) and (\ref{eq.37}). The first system reflects the field equation in $f(Q)$ gravity under $T_{i\,j}$, whereas the second system represents the additional source $\theta_{i\,j}$. As a consequence, it is assumed that the energy-momentum tensor ${T}_{i\,j}$ describes an anisotropic matter distribution with,
    \begin{eqnarray}
    && \hspace{-3cm} T_{i\,j} = \left(\rho+p_t\right)u_{i}\,u_{j} + p_t\,g_{i\,j} + \left(p_r-p_t\right) v_{i}\,v_{j}, \label{eq.38}
    \end{eqnarray}
where $\rho$ stands for the energy density while the four-velocity vector and unitary space-like vectors are represented by $u_i$ and $v_i$, respectively. Furthermore, they both fulfill the requirement $u_i u^j = -v_i v^j = -1$. The radial and tangential pressures, denoted by $\rho_r$ and $p_t$, are both functions of the radial coordinate $r$ such that,
    \begin{eqnarray} 
        && \hspace{-2.1cm}\rho^{\text{eff}}=\rho+\alpha \,\theta^0_0,~~p^{\text{eff}}_r=p_r-\alpha\,\theta^1_1,~~p^{\text{eff}}_t=p_t-\alpha\,\theta^2_2,~~~~ \label{eq.39}
    \end{eqnarray}
and the corresponding effective anisotropy,
    \begin{eqnarray} 
        && \hspace{-5.5cm}\Delta^{\text{eff}}=p^{\text{eff}}_t-p^{\text{eff}}_r= \Delta+\Delta_{\theta}, \label{eq.40}
    \end{eqnarray}
where $~\Delta= p_t-p_r~~\text{and}~~\Delta_\theta= \alpha (\theta^1_1-\theta^2_2)\nonumber.$\\ 
It should be remembered that $T_{i\,j}$ and $\theta_{i\,j}$ are two anisotropies that make up effective anisotropy. Gravitational decoupling produces the anisotropy $\Delta_{\theta}$, which modifies the effective anisotropy, but this transformation is entirely dependent on the behavior of $\Delta_{\theta}$. The following set of equations of motion dependent on the gravitational potentials $W$ and $H$, or when $\beta = 0$, are produced by putting equations (\ref{eq.36}) and (\ref{eq.37}) into the system (\ref{eq.32})-(\ref{eq.34}):
    \begin{eqnarray}
        && \hspace{-0.8cm}\rho+ \frac{q^2}{r^4}= \frac{\beta_1 (1-W) }{r^2}-\frac{W^{\prime} \beta_1 }{r}-\frac{\beta_2 }{2},\label{eq.41}\\
        && \hspace{-0.8cm} p_r-\frac{q^2}{r^4}=\frac{\beta_1 (W-1) }{r^2}+\frac{H^{\prime} W \beta_1 }{r}+\frac{\beta_2 }{2}, \label{eq.42}\\
        && \hspace{-0.8cm} p_t+\frac{q^2}{r^4}=\frac{\beta_1(W^{\prime} H^{\prime} + 2 H^{\prime \prime} W +H^{\prime 2} W)  }{4} +\frac{\beta_1\,(W^{\prime} +H^{\prime} W) }{2 r}+\frac{\beta_2}{2}, \label{eq.43}\nonumber \\
    \end{eqnarray}
and according to the TOV equation (\ref{eq.35}),
    \begin{eqnarray}
        && \hspace{-2.8cm}-\frac{H^\prime}{2}(\rho+p_r)-(p_r)^{\prime}+\frac{2}{r}( p_{t}-p_r)+\frac{2qq'}{r^4}=0.~~\label{eq.44}
    \end{eqnarray}
Consequently, the space-time that follows can provide the corresponding solution:
    \begin{eqnarray}
        && \hspace{-2.9cm}ds^2=-e^{H(r)}dt^2+\frac{dr^2}{W(r)}+r^2d\theta^2+r^2\text{sin}^2\theta d\phi^2, \label{eq.45}
    \end{eqnarray}
Moreover, the system of field equations for $\theta$-sector is derived by turning on $\beta$ as,
    \begin{eqnarray}
        && \hspace{-1.4cm}\theta^{0}_0=-\beta_1 \Big(\frac{\Psi   }{r^2}+\frac{\Psi^\prime }{r}\Big), \label{eq.46}\\
        && \hspace{-1.4cm}\theta^1_1=-\beta_1\Big[\frac{\Psi  }{r^2}+\frac{(\Phi^{\prime} \Psi +W\,\xi^{\prime})  }{r}\Big], \label{eq.47}\\
        && \hspace{-1.4cm}\theta^2_2=-\beta_1 \Big[\frac{(\Psi^\prime \Phi^{\prime} + 2 \Phi^{\prime \prime} \Psi+\Phi^{\prime 2} \Psi +W^{\prime}\,\xi^{\prime}) }{4} +\frac{(\Psi^\prime +\Phi^{\prime} \Psi) }{2 r}\Big]\nonumber\\&&\hspace{-0.7cm} -\beta_1\Big[\frac{W}{4}\,\big(2\,\xi^{\prime\prime}+\beta_1\,\xi^{{\prime}\,2}+\frac{2\,\xi^{\prime}}{r}+2\,H^{\prime}\,\xi^{\prime}\big)\Big],~~~~~~ \label{eq.48}
    \end{eqnarray}
and the associated conservation is,
    \begin{eqnarray}
    && \hspace{-2.6cm}-\frac{H^{\prime}}{2} (\theta^0_0-\theta^1_1)+ (\theta^1_1)^\prime+\frac{2}{r} (\theta^1_1-\theta^2_2)=\frac{\xi^{\prime}}{2}\,({p_r}+{\rho}). \label{eq.49}
    \end{eqnarray}
However, the mass function for both systems is given by
    \begin{eqnarray}\label{eq.50}
        && \hspace{-4.4cm}m_{Q}=\frac{1}{2} \int^r_0 \left(\rho(x)+\frac{q^2}{x^2}\right)\, x^2 dx+\frac{q^2}{2r} \nonumber \\&&\hspace{-4.4cm}\text{and}~~m_{\theta}= \frac{1}{2}\,\int_0^r \theta^0_0 (x)\, x^2 dx, ~~~
    \end{eqnarray}
where the relevant mass functions for the sources $T_{ij}$ and $\theta_{ij}$ are $m_{Q}(r)$ and $m_{\theta}(r)$, respectively. Then, in the context of $f(Q)$ gravity, the interior mass function of minimally deformed space-time (\ref{eq.17}) may be expressed as,
    \begin{eqnarray} \label{eq.51}
        && \hspace{-5.1cm}\hat{m}_Q(r)=m_Q(r)-\frac{\beta_1\,\alpha}{2}\,r\,\Psi(r).
    \end{eqnarray}
Before moving on to the discussion of the solution, we discussed the suitable boundary condition to derive the arbitrary constants for self-bound compact objects in the next section.   

\section{Boundary conditions} \label{sec3}
The boundary conditions play a crucial role in the exploration of the compact star. Finding a vacuum solution, or external space-time, that is compatible with internal space-time at the pressure-free boundary at $r=R$ is essential for this study. The Reissner–Nordström–(anti)-de Sitter (RNdS) space-time is the most suitable exterior space-time for spherically symmetric charged compact objects in $f(Q)$ gravity that can be given as 
    \begin{eqnarray} \label{eq.52}
        && \hspace{-1cm}ds^2_+ =-\bigg(1-\frac{2{\mathcal{M}}}{r}+\frac{\tilde{\mathcal{Q}}^2}{r^2}-{\frac{\Lambda}{3}}~r^2\bigg)\,dt^2+\frac{dr^2}{\bigg(1-\frac{2{\mathcal{M}}}{r}+\frac{\tilde{\mathcal{Q}}^2}{r^2}-{\frac{\Lambda} {3}}~r^2\bigg)} \nonumber \\ && \hspace{-0.2cm} +r^2 \Big(d\theta^2  +\sin^2\theta\,d\phi^2 \Big),\label{metric1}~~~~~~
    \end{eqnarray} 
where $\mathcal{M}$ and $\tilde{\mathcal{Q}}$ are the total mass and total electric charge, respectively.  While $\Lambda$ denotes a cosmological constant. Then $\mathcal{M}=\hat{M}_Q/\beta_1$, and $\Lambda=\beta_2/2\beta_1$, where $\hat{m}_Q(R)=\hat{M}_Q$. It is clearly observed that when $\beta_1=1$ and $\beta_2=0$, the RNdS space-time (\ref{eq.53}) reduces into the RN exterior solution. On the other hand, the gravitationally deformed interior space-time for the region ($0 \le r \le R$)  is given by  
    \begin{eqnarray}\label{eq.53}
        && \hspace{-1.6cm}ds^2_{-}= -e^{H(r)+\alpha\,\xi(r)}\,dt^2+ \big[W(r)+\alpha\,\Psi(r)\big]^{-1} dr^2 +r^2 d\theta^2 \nonumber\\&& \hspace{-0.7cm} +r^2 \sin^2\theta\,d\phi^2, \label{metric2}~~~~~ 
    \end{eqnarray} 
Israel-Darmois matching criteria require that the solution meet the first and second form at the boundary surface of the star ($r=R$) in order to provide a smooth matching between the internal and external spacetimes.
Mathematically, it can be written as  
    \begin{eqnarray}
        \label{eq.54}
        e^{\Phi^{-}(r)}|_{r=R}=e^{\Phi^{+}(r)}|_{r=R}~~~\mbox{and}~~~~e^{\lambda^{-}(r)}|_{r=R}=e^{\lambda^{+}(r)}(r)|_{r=R}, \nonumber \\
    \end{eqnarray}
and 
    \begin{eqnarray}
        &&\hspace{-2.8cm} \big[G_{i\,\varepsilon}\, r^{\varepsilon}\big]_{\Sigma} \equiv \lim_{r \rightarrow R^{+}} (G_{\epsilon\,\varepsilon})-\lim_{r \rightarrow R^{-}} (G_{\epsilon\,\varepsilon})=0~~ \nonumber\\&& \hspace{-2.8cm}\Longrightarrow~~~ \big[T^{\text{eff}}_{\epsilon\,\varepsilon}\,r^{\varepsilon}\big]_{\Sigma}=\big[(T_{\epsilon\,\varepsilon}+\alpha\,\theta_{\epsilon\,\varepsilon})\,r^{\varepsilon}\big]_{\Sigma}=0,~~~ \label{eq.55}
    \end{eqnarray}
The conditions (\ref{eq.54}) and (\ref{eq.55}) yields,
    \begin{eqnarray}
        && \hspace{-1.6cm}e^{H(R)+\alpha \xi(R)} = \bigg(1-\frac{2{\mathcal{M}}}{R}+\frac{\tilde{\mathcal{Q}}^2}{R^2}-{\frac{\Lambda}{3}}~R^2\bigg) \label{eq.56}\\
        && \hspace{-1.6cm}W(R)+\alpha\,\psi(R) = \bigg(1-\frac{2{\mathcal{M}}}{R}+\frac{\tilde{\mathcal{Q}}^2}{R^2}-{\frac{\Lambda}{3}}~R^2\bigg), \label{eq.57}\\
        && \hspace{-1.6cm}p^{\text{eff}}_r(R) = p_r(R)+\alpha\,\alpha_1\, \,\Big[\psi_{_\Sigma}\Big(\frac{1 }{R^2}+\frac{H^{\prime}_{_\Sigma}   }{R}\Big)+\frac{W_{_\Sigma}\,\xi^{\prime}_{_\Sigma}}{R} \Big]=0. ~~~~~\label{eq.58}
    \end{eqnarray}

Using the equations (\ref{eq.56}) -- (\ref{eq.58}), we have determined the unknown parameters for both solutions, such as Bag constant ($\mathcal{B}_g$), mass ($\mathcal{M}$) and arbitrary constant ($C$) to calculate the numerical value.

\section{Gravitationally decoupled solution in \texorpdfstring{$\lowercase{f}(Q)$}{} gravity} \label{sec4}

Accordingly, the MIT bag model is the most elementary phenomenological description of quark matter. In terms of quarks, such models have been developed to describe the properties of hadrons. The range in which quarks are contained is known as a bag, and the energy required to produce it per unit volume is known as a bag pressure, $\mathcal{B}_g$. Further, the quarks cannot reach outside the bag because they are free inside. The first step towards solving the structure equations is identifying the sources. In addition, we consider an extension of the MIT bag model \citep{Chodos_1974_9, Chodos_1974_10, Farhi_1984_30} in which matter inside the star can be modelled as a relativistic gas of de-confined quarks. To solve the first system, we use the equation of the MIT bag model equation of state,
    \begin{eqnarray}
        && \hspace{-6.8cm}p_r=\frac{1}{3}(\rho-4 \mathcal{B}_g) \label{eq59}
    \end{eqnarray}
where parameter $\mathcal{B}_g$ represents the bag constant, while $p_r$ and $\rho$ represent pressure and energy density. It must be noted that when $\rho = 4 \mathcal{B}_g$, the external pressure acting on a bag filled with quarks vanishes. Bag constant $\mathcal{B}_g$ has been measured to have a valid range of $57\leq \mathcal{B}_g \leq 92$ MeV/fm$^3$ \citep{Fiorella_2018_457, Blaschke_2018_457}. 

By using the above EOS of state, we get the following differential equations, 
    \begin{small}
    \begin{eqnarray}
        4 \mathcal{B}_g r^2+\beta_1\,r W'+3 \beta_1 r\,H'W+4 \beta_1 (W-1)+2 \beta_2 r^2+4 E^2 r^2  =0, \label{eq:diff_eos}
    \end{eqnarray}
    \end{small}
Using the Buchdahl model to solve the above differential equation (\ref{eq:diff_eos}), we assume two well-known ansatzes for $W$, $H$ and $E$. We can solve the differential equation containing $W(r)$ using a well-known Buchdahl ansatz with these assumptions,
    \begin{eqnarray}
        && \hspace{-6.5cm}W(r)= \frac{K+Cr^2}{K (1+Cr^2)}, \label{eq61}
    \end{eqnarray}
and electric field of the form,
    \begin{eqnarray}
        && \hspace{-7cm}E^2=\frac{E_0 C^2 r^2}{(1+Cr^2)^2}, \label{eq62}
    \end{eqnarray}
where $E_0$ is a constant. This equation (\ref{eq61}) defines the geometry of the star by using $K$ and $C$ parameters. Note that \citet{Buchdahl_1959_116} proposed the ansatz for the metric function $g_{rr}$ to develop a realistic model for a relativistic compact star.  Many authors have pointed out that this metric potential generates a viable stellar models which has a non-singular energy density that decreases outward. The metric function equation (\ref{eq61}) is positive, free from the singularity at $r=0$, and monotonically increasing outward in addition to the above. In this section, we will show how an analytical Buchdahl model can be expanded to account for both positive and negative values of the spheroidal parameter $K$. Depending on the two factors, the energy density or pressure will be negative in the analysis that follows over the range of $0< K< 1$. When $K=0$, one can obtain the Schwarzschild interior solution, while at $K=1$, the hypersurfaces $\{t = \mbox{constant} \}$ are flat. More generally, one could retrieve the \citet{Vaidya_1982_3} solution when $C = -K/R^2$ and \citet{Durgapal_1983_27} solution when $K=-2$. Moreover, there is a solution for charged and uncharged perfect fluid has been also examined \citep{Gupta_2005_299, Gupta_2003_283}. In our current study, we perform our analysis to check the validity of the solution by taking $K=-2$. 

By substituting the $W(r)$ and $E(r)$ in the equation (\ref{eq61}) and integration,  we get the following solution for $H(r)$, 
    \begin{eqnarray}
        && \hspace{-1.2cm}H(r) =\frac{1}{3 \beta_1 C (K-1)}\Big[\log \left(-C r^2-K\right) \big\{2 \mathcal{B}_g K (K-1)^2\nonumber\\&& \hspace{0.1cm}-2 C E_0 K^2+\beta_1 C \left(2 K^2-5 K+3\right)+\beta_2 K (K-1)^2\big\}\nonumber\\&& \hspace{0.1cm}+(K-1) K (-2 \mathcal{B}_g-\beta_2) \left(C r^2+1\right)+C \log \left(C r^2+1\right) \nonumber\\&& \hspace{0.1cm} \times (2 E_0 K+\beta_1 (K-1))\Big]+F,~~~~~~\label{eq63}
    \end{eqnarray}

Eqs. (\ref{eq61})--(\ref{eq63}) show our space-time geometry for the seed solution. The solution of the second system of equations (\ref{eq.46})--(\ref{eq.48}) is, however, necessary for the $\theta$-sector. This situation, therefore, requires that two additional pieces of information be provided to close the theta-system, such as $\Psi$ and $\xi(r)$, because there are three independent equations with five unknowns. To simplify, we assume $\xi(r)=H(r)$, which provides $\Phi(r)=(1+\alpha) H (r)$. To maintain physical viability, $\Phi (r)$ should have a monotonic increase towards the boundary, so $H(r)+\alpha \xi (r)$ must also have an increasing function of $r$. As discussed in the following reference, we solve the system of equations (\ref{eq.46})--(\ref{eq.48}) with two mimic approaches: (i) $\theta^0_0$ with energy density $\rho$, i.e. $\rho=\theta^0_0$, and (ii) $\theta^1_1$ with radial pressure $p_r$, i.e. $p_r=\theta^1_1$ [Ref.~\citep{Ovalle_2017_95}]. These two methods provide the following two equations:
    \begin{eqnarray}
        &&\hspace{-1.6cm} \Psi(r)= \frac{6\beta_1 (W(r)-1)+\beta_2 r^3}{6 \beta_1}~~\label{eq64}\\
        &&\hspace{-1.6cm} \Psi(r)=\frac{2 \beta_1 \big(1-W(r) [1+r \{H'(r)+\xi'(r)\big\}]\big)-\beta_2 r^2}{ 2 \beta_1[1+r\,\Phi'(r)]}.~~~~~~~\label{eq65}
    \end{eqnarray}
As shown in Ref. \citep{Sharif_2020_365, Sharif_2020_29, Maurya_2020_30, Maurya_2021_81, Maurya_2022_925, Maurya_2022_519}, $\Psi(0)=0$ and $\Psi(r)$ is free from any singularity. This technique has been successful applications in modelling compact objects in GR, the modification of gravity theories in Ref. \citep{Contreras_2022_82} and the gravitational cracking concept under gravitational decoupling. We used these methods to solve the second system because these works inspired us.

\subsection{Mimicking to density approach \texorpdfstring{$\rho=\theta^0_0$}{}} \label{solA}

By mimicking of the density approach gives the following differential equation,
    \begin{eqnarray}\label{eq60}
        && \hspace{-3cm} 2\beta_1 (r\,\Psi^{\prime} 
        +\Psi)+2\beta_1(1-W-r\,W^{\prime})  - r^2\,\beta_2 =0.
    \end{eqnarray}
On inserting the metric function $W(r)$ and integrating the above equation, we get 
    \begin{eqnarray}
        &&\hspace{-2cm}\Psi(r)=\frac{r^2 \left[C \left(6 \beta_1-3 E_0 K-6 \beta_1 K+\beta_2 K r^2\right)+\beta_2 K\right]}{6 \beta_1 K \left(C r^2+1\right)},
    \end{eqnarray}
Now inserting expressions $\Psi(r)$ and [$\xi(r)=H(r)$] along with metric function in the second system, we can get the expressions for $\theta^0_0$, $\theta^1_1$, and $\theta^2_2$. In this way, we obtain the expressions for effective quantities as 
    \begin{small}
    \begin{eqnarray}
        && \hspace{-0.6cm} \rho^{\text{eff}}=\frac{(1+\alpha)}{2 K \left(C r^2+1\right)^2}\Big[6 \beta_1 C (K-1)-C^2 r^2 \big[2 E_0 K-2 \beta_1 (K-1)\nonumber\\
        && \hspace{0.5cm}+\beta_2 K r^2\big]-2 \beta_2 C K r^2-\beta_2 K\Big], \\
        &&\hspace{-0.6cm} p^{\text{eff}}_r=\frac {1}{6 K \left(C r^2+1\right)^2}\Big[C^2 r^2 \Big(-3 E_0 K \big[\alpha +\alpha ^2 N_{11}(r) r^2+\alpha  N_{11}(r) r^2\nonumber\\
        && \hspace{-0.1cm} -2\big]+\beta_2 K r^2 \big[\alpha+\alpha ^2 N_{11}(r) r^2+\alpha  N_{11}(r) r^2+3\big]-6 (\alpha +1) \beta_1 \nonumber\\
        && \hspace{0.5cm} \times \left(\alpha  N_{11}(r) K r^2-(\alpha +1) N_{11}(r) r^2+K-1\right)\Big) \nonumber\\
        && \hspace{-0.1cm} +C \Big\{-3 \alpha  E_0 K \left((\alpha +1) N_{11}(r) r^2+1\right)+2 \beta_2 K r^2 \big(\alpha +\alpha ^2 N_{11}(r) r^2\nonumber\\
        && \hspace{0.5cm}+\alpha  N_{11}(r) r^2+3\big)-6 (\alpha +1) \beta_1 \big[(\alpha -1) N_{11}(r) K r^2-\nonumber\\
        && \hspace{0.5cm} (\alpha +1) N_{11}(r) r^2+K-1\big]\Big\}+K \Big\{(\alpha +3) \beta_2+(\alpha +1) \nonumber\\
        && \hspace{0.5cm} \times N_{11}(r) \left(6 \beta_1+\alpha  \beta_2 r^2\right)\Big\}\Big],\\
        && \hspace{-0.6cm} p^{\text{eff}}_t=\frac{1}{24 K \left(C r^2+1\right)^2}\Big[C^2 \Big\{r^4 \big[-2 (\alpha +1) N_{12}(r) \big(6 \beta_1 (\alpha  (K-1)-1)\nonumber\\
        && \hspace{0.5cm}-\alpha  \beta_2 K r^2\big)+(\alpha +1)^2 \times N_{11}(r)^2 r^2 \big(\beta_1 (6-6 \alpha  (K-1))\nonumber\\
        && \hspace{-0.1cm}+\alpha  \beta_2 K r^2\big)-4 (\alpha +1) N_{11}(r) \left(3 \beta_1 (\alpha  (K-1)-1)-\alpha  \beta_2 K r^2\right)\nonumber\\
        && \hspace{0.5cm} +4 (\alpha +3) \beta_2 K\big]-3 E_0 K r^2 \big(2 \alpha  (\alpha +1) N_{12}(r) r^2+\alpha  (\alpha +1)^2 \nonumber\\
        && \hspace{0.5cm}\times N_{11}(r)^2 r^4+2 \alpha  (\alpha +1) N_{11}(r) r^2+8\big)\Big\}+N_{33}(r)\Big], 
    \end{eqnarray}

\subsection{Mimicking to pressure constraints approach \texorpdfstring{$p_r=\theta^1_1$}{}} \label{solB}

By using mimicking of radial pressure, we can get the direct expression for deformation function $\Psi(r)$ as  
    \begin{eqnarray}
        &&\hspace{-0.75cm} \Psi(r)= -r^2\,\Big[\left(C r^2+K\right) \Big\{16 \mathcal{B}_g K \left(C r^2+1\right)^2+5 C^2 r^2 \big(2 \beta_1\nonumber\\
        && \hspace{0.5cm}+2 E_0 K-2 \beta_1 K+\beta_2 K r^2\big)-18 \beta_1 C \times (K-1)+10 \beta_2 C K r^2\nonumber\\
        && \hspace{-0.1cm}+5 \beta_2 K\Big\}\Big] \Big/ \Big[2 K \left(C r^2+1\right) \Big\{4 (\alpha +1) \mathcal{B}_g K \left(C r^3+r\right)^2+C^2 r^4 \nonumber\\
        && \hspace{-0.1cm} \times \left(4 \alpha  \beta_1+\beta_1+4 (\alpha +1) E_0 K-4 (\alpha +1) \beta_1 K+2 (\alpha +1) \beta_2 K r^2\right)\nonumber\\
        && \hspace{0.5cm}+C r^2 \big(\beta_1 (6 \alpha -3 (2 \alpha +3) \times K+3)+4 (\alpha +1) \beta_2 K r^2\big)\nonumber\\
        && \hspace{0.5cm}+K \left(2 (\alpha +1) \beta_2 r^2-3 \beta_1\right)\Big\}\Big],
    \end{eqnarray}
Again by substituting the deformation functions along with the metric function in the second system, we can get the expression for effective quantity $\rho^{\text{eff}}$, $p^{\text{eff}}_r$, and  $p^{\text{eff}}_t$ as,
    \begin{eqnarray}
        &&\hspace{-0.9cm}  \rho^{\text{eff}}= \alpha  \beta_1 \Bigg[\frac{1}{N_{13}(r)}\Big\{\left(C r^2+K\right) \Big(16 \mathcal{B}_g K \left(C r^2+1\right)^2+5 C^2 r^2 \nonumber\\
        && \hspace{0.5cm} \times\left(2 \beta_1+2 E_0 K-2 \beta_1 K+\beta_2 K r^2\right) -18 \beta_1 C (K-1)\nonumber\\
        && \hspace{0.5cm}+10 \beta_2 C K r^2+5 \beta_2 K\Big)\Big\}-N_{11}(r)\Bigg]-\frac{1}{2 K \left(C r^2+1\right)^2}\nonumber\\
        && \hspace{-0.1cm} \times\Big[C^2 r^2 \big(2 E_0 K-2 \beta_1 (K-1)+\beta_2 K r^2\big)-6 \beta_1 C (K-1)\nonumber\\
        && \hspace{0.5cm}+2 \beta_2 C K r^2+\beta_2 K\Big], \\
        &&\hspace{-0.9cm} p^{\text{eff}}_r=\frac{1}{6 K \left(C r^2+1\right)^2}\Big[(\alpha -1) \Big\{8 \mathcal{B}_g K \left(C r^2+1\right)^2+C^2 r^2 \nonumber\\
        && \hspace{0.5cm} \times\left(2 \beta_1+2 E_0 K-2 \beta_1 K+\beta_2 K r^2\right)+C \big(6 \beta_1-6 \beta_1 K\nonumber\\
        && \hspace{0.5cm} +2 \beta_2 K r^2\big)+\beta_2 K\Big\}\Big], \\
        && \hspace{-0.9cm} p^{\text{eff}}_t=\frac{1}{4} \Bigg[\frac{1}{K \left(C r^2+1\right)^2}\Big\{C^2 \big(r^4 \Big(2 \beta_1 N_{12}(r)+\beta_1 N_{11}(r)^2 r^2 \nonumber\\
        && \hspace{0.5cm}+2 \beta_1 N_{11}(r)+2 \beta_2 K\big)-4 E_0 K r^2\Big)+C\Big[\beta_1 \Big(K \big(2 N_{12}(r) r^2 \nonumber\\
        && \hspace{0.5cm}+N_{11}(r)^2 r^4-4\big)+2 N_{12}(r) r^2+\left(N_{11}(r) r^2+2\right)^2\Big)\nonumber\\
        && \hspace{0.5cm}+4 \beta_2 K r^2\Big]+K \big(2 \beta_2+2 \beta_1 N_{12}(r)+\beta_1 N_{11}(r)^2 r^2\nonumber\\
        && \hspace{0.5cm}+2 \beta_1 N_{11}(r)\big)\Big\}+\frac{N_{20}(r)}{K \left(C r^2+1\right)^2}+\alpha  \beta_1 N_{11}(r) \big[(\alpha +1) \nonumber\\
        && \hspace{0.5cm} \times  N_{11}(r) r^2+2\big]-\frac{N_{22}(r)}{N_{21}(r)}\Bigg]
    \end{eqnarray}
    \end{small}

The expressions for used notations are given in the Appendix.

    \begin{figure*}
        \centering
        \includegraphics[width=8cm,height=6.5cm]{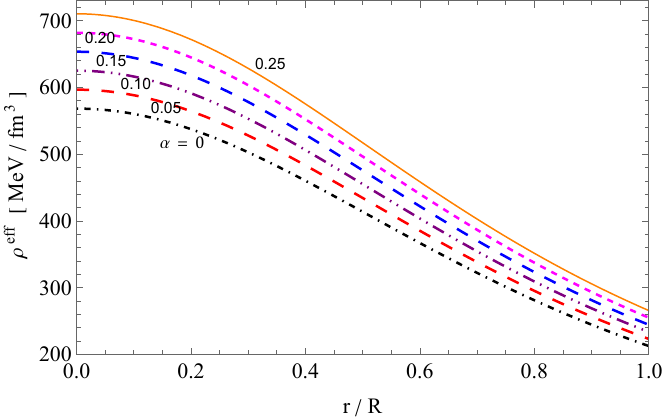}~~ \includegraphics[width=8cm,height=6.5cm]{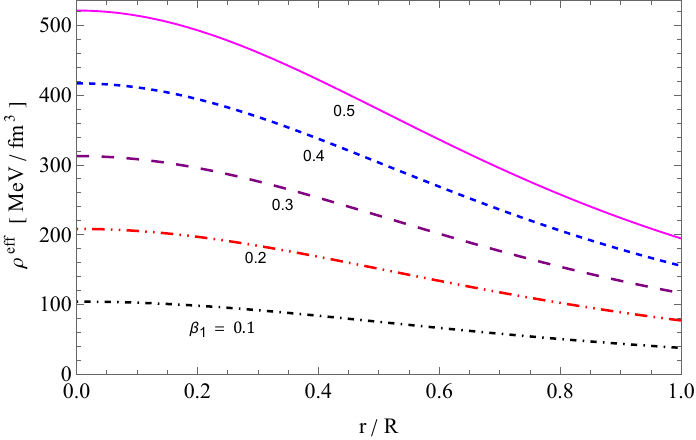}
        \caption{The behavior of effective  energy density ($\rho^{\text{eff}}$ in $\text{MeV}/\text{fm}^3$ [$1\,\text{MeV}/\text{fm}^3=1.3234 \times 10^{-6}~ \text{km}^{-2}$]) for different values of constants $\alpha$-(left panel) and $\beta_1$ -(right panel) for the solution \ref{solA} ($\rho=\theta^0_0$). The following numerical values are employed to plot the figures: $C= 0.007/km^2$, $K = -2$, $\beta_1 = 0.6$, $\beta_2 = 0$ and $E_0 = 0.02$ (left Fig) and $C= 0.007/km^2$, $K = -2$, $\alpha = 0.1$, $\beta_2 = 0$ and $E_0 = 0.02$ (right Fig).  } \label{fig: rho0}
    \end{figure*}
    \begin{figure*} 
        \centering
        \includegraphics[width=8cm,height=6.5cm]{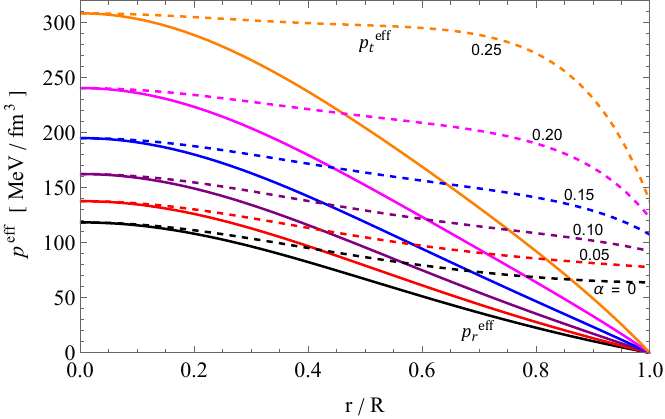}~~~\includegraphics[width=8cm,height=6.5cm]{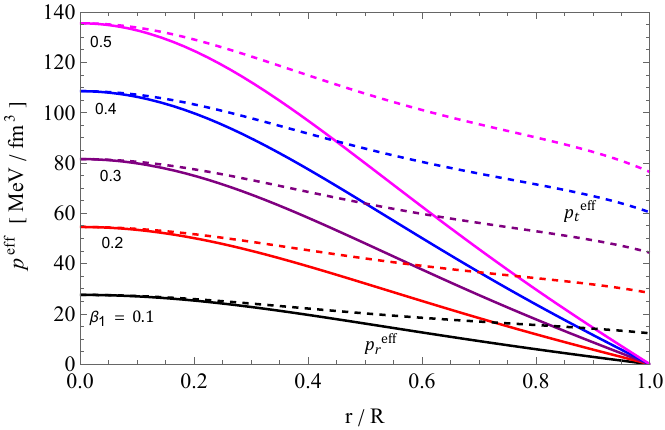}
        \caption{The behavior of effective radial and tangential pressures ($p^{\text{eff}}_r$ \& $p^{\text{eff}}_t$ in $\text{MeV}/\text{fm}^3$) for different values of constants $\alpha$-(left panel) and $\beta_1$ -(right panel) for the solution \ref{solA} ($\rho=\theta^0_0$). The following numerical values are employed in Fig. \ref{fig: rho0}.} \label{fig: press0}
    \end{figure*} 

    \begin{figure*} 
        \centering
        \includegraphics[width=8cm,height=6.5cm]{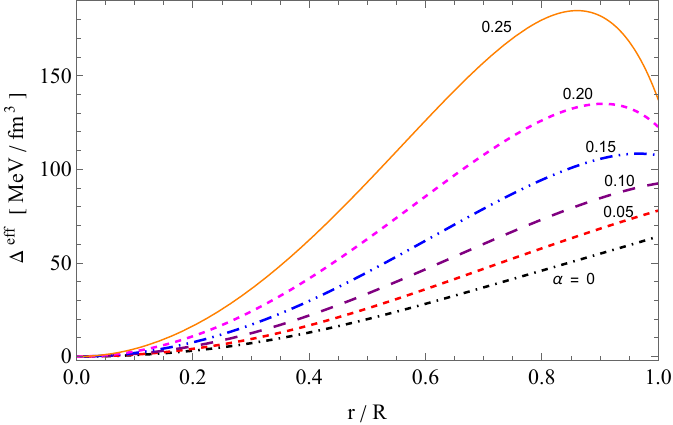}~~~\includegraphics[width=8cm,height=6.5cm]{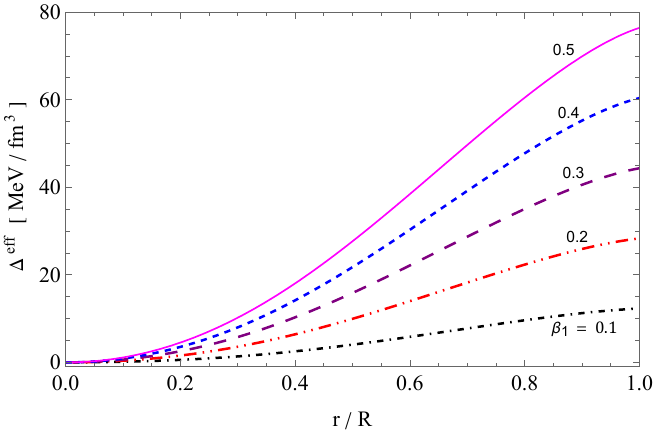}
        \caption{The behavior of effective anisotropy ($\Delta^{\text{eff}}$ in $\text{MeV}/\text{fm}^3$ ) for different values of constants $\alpha$-(left panel) and $\beta_1$ -(right panel) for the solution \ref{solA} ($\rho=\theta^0_0$). The following numerical values are employed in Fig. \ref{fig: rho0}}. \label{fig: anisotropy0}
    \end{figure*}

    \begin{figure*} 
        \centering
        \includegraphics[width=8cm,height=6.5cm]{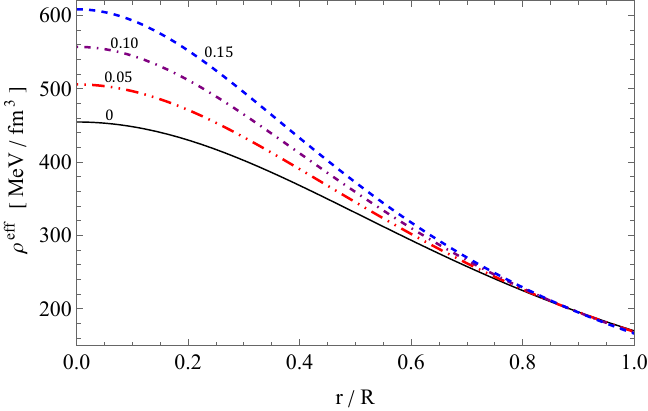}~~ \includegraphics[width=8cm,height=6.5cm]{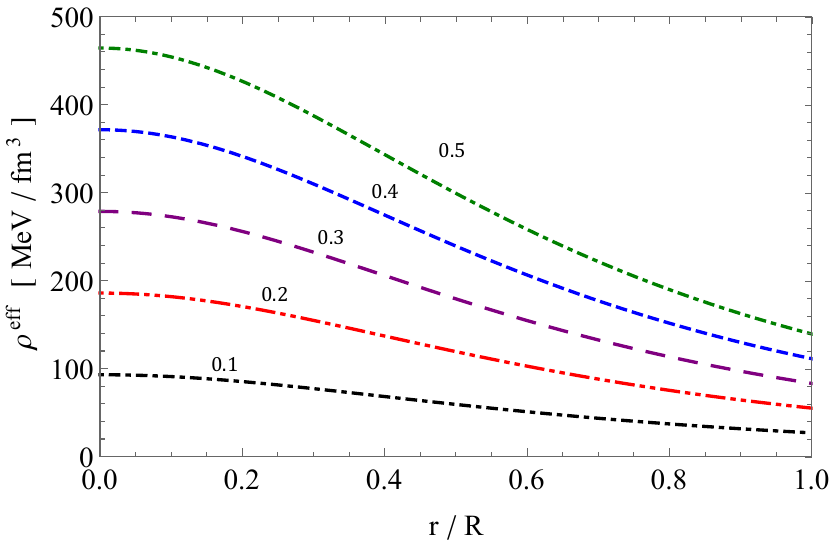}
        \caption{The behavior of effective  energy density ($\rho^{\text{eff}}$ in $\text{MeV}/\text{fm}^3$) for different values of constants $\alpha$-(left panel) and $\beta_1$ -(right panel) for the solution \ref{solB} ($p_r=\theta^1_1$). The following numerical values are employed to plot the figures:   $C= 0.007/km^2$, $K = -5$, $\beta_1 = 0.6$, $\beta_2 = 0$ and $E_0 = 0.02$ (left Fig) and $C= 0.007/km^2$, $K = -5$, $\alpha = 0.1$, $\beta_2 = 0$ and $E_0 = 0.02$ (right Fig). } \label{fig: rho1}
    \end{figure*}
    \begin{figure*}
        \centering
        \includegraphics[width=8cm,height=6.5cm]{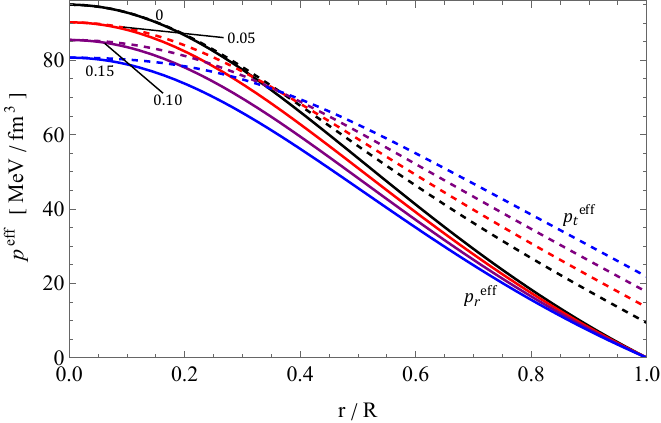}~~~\includegraphics[width=8cm,height=6.5cm]{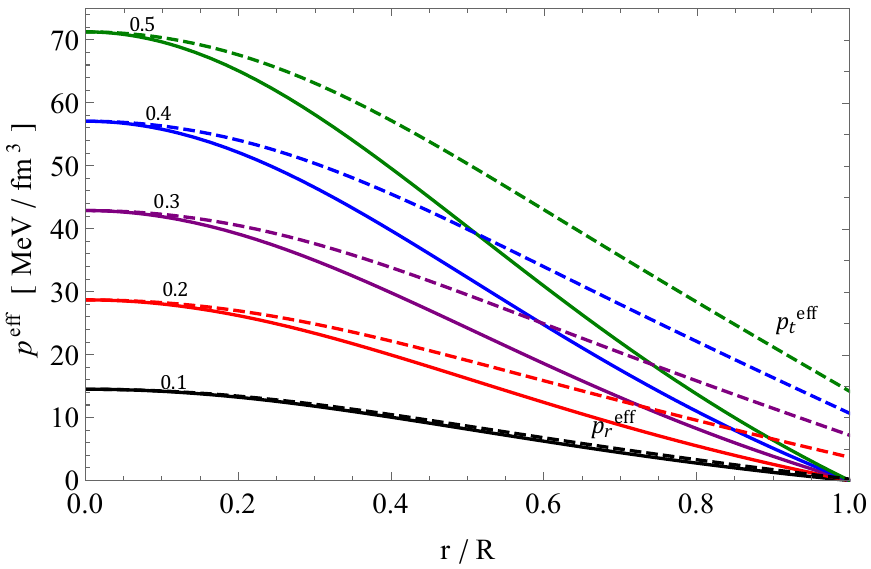}
        \caption{The behavior of effective radial and tangential pressures ($p^{\text{eff}}_r$ \& $p^{\text{eff}}_t$ in $\text{MeV}/\text{fm}^3$) for different values of constants $\alpha$-(left panel) and $\beta_1$ -(right panel) for the solution \ref{solB} ($p_r=\theta^1_1$). The following numerical values are employed in Fig. \ref{fig: rho1}.} \label{fig: press1}
    \end{figure*}

    \begin{figure*}
        \centering
        \includegraphics[width=8cm,height=6.5cm]{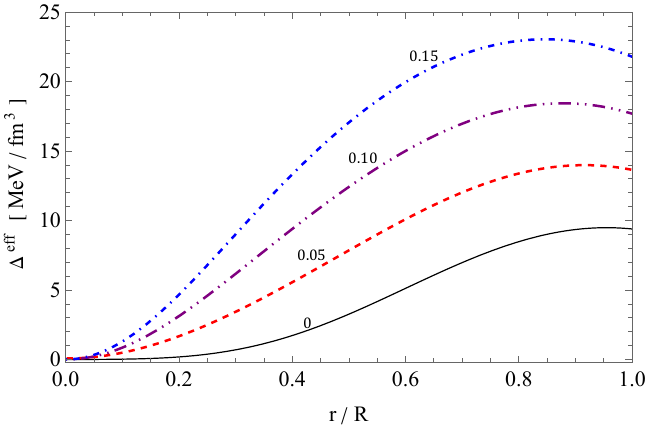}~~~\includegraphics[width=8cm,height=6.5cm]{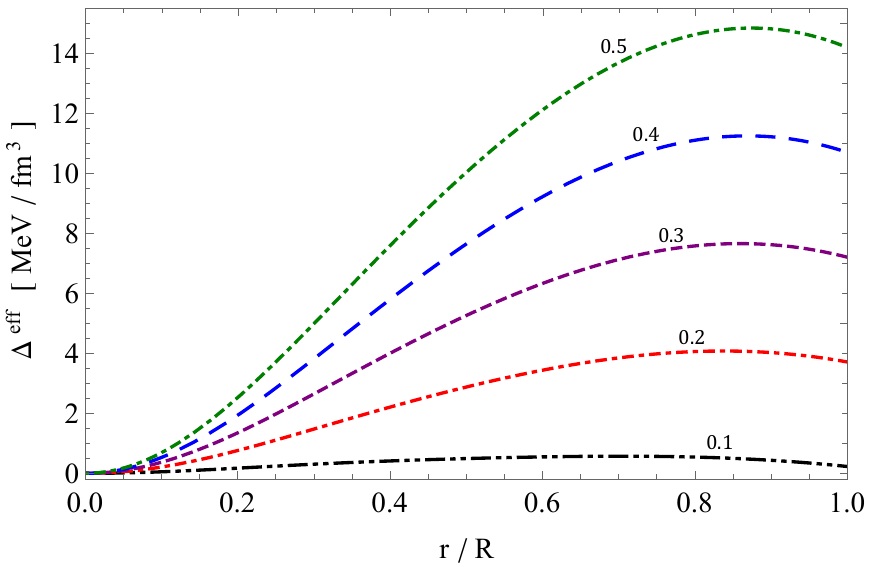}
        \caption{The behavior of effective anisotropy ($\Delta^{\text{eff}}$ in $\text{MeV}/\text{fm}^3$) for different values of constants $\alpha$-(left panel) and $\beta_1$ -(right panel) for the solution \ref{solB} ($p_r=\theta^1_1$). The following numerical values are employed in Fig. \ref{fig: rho1}.} \label{fig: anisotropy1}
    \end{figure*}

\section{Physical Analysis and Astrophysical implications } \label{sec5}

\subsection{Regular SS model behavior with minimum deformation}
We present a detailed physical analysis of our solutions based on the graphical plots presented here, and we present a detailed physical analysis of our solutions. In order to begin, we need to analyze the $\theta_0^0 =\rho$ solution. According to Fig. \ref{fig: rho0}, the density decreases monotonically with the scaled radial coordinate, $r/R$. The deformation parameter, $\beta_1$, has been fixed, and this is represented by the $Q$ switch of $\alpha$, which we have varied. Accordingly, it is clear that as $\alpha$ increases, the density of the compact object also increases, as it increases with $\alpha$. Fig. \ref{fig: press0} shows the results of keeping $\alpha$ constant throughout and varying the deformation parameter in the bottom right panel. This scenario shows a minimal variation in effective density with increasing $\beta_1$. Fig. \ref{fig: press0} shows the effective radial pressure at each interior point of the stellar configuration. The pressure gradually decreases as one moves from the centre to the boundary. Fig. \ref{fig: press0} shows that the effective radial pressure vanishes when we reach a boundary because the energy flux into exterior space-time is no longer present. As $\alpha$ increases, we observe that the effective radial pressure increases proportionately to the nonmetricity scalar, $Q$. Fig. \ref{fig: press0} shows the radial and transverse stress resulting from increasing the decoupling parameter. A similar observation also holds for effective tangential pressure. There are only very slight deviations in both the effective radial and the effective transverse pressure. Also, the effective tangential pressure dominates the radial one in the surface layers.

As shown in Fig. \ref{fig: anisotropy0}, the effective anisotropy parameter is plotted with the non-metricity scalar $Q$ switch variable while the decoupling constant remains constant. When the $Q$ switch is varied, the anisotropy parameter is negative up to a certain radius, $r_0$. As a result, an attractive force arises due to anisotropy when the effective radial pressure dominates its tangential counterpart. The anisotropy of the surface layers of the star becomes positive as one moves away from $r = r_0$ to the boundary, indicating a repulsive force that stabilizes them. When the decoupling constant is varied in the top right panel, $\Delta^\text{eff}$ displays an anisotropy parameter that decreases with an increase in $\beta_1$, indicating that the decoupling constant quenches contributions from anisotropy. The anisotropy of the central regions of the star is also negative, indicating that these regions are unstable as opposed to the repulsive contributions from $\Delta^\text{eff}$.

Figs. \ref{fig: rho1} and \ref{fig: press1} present the effective stresses and densities for $\alpha$ varying and $\beta_1$ fixed, and another one with $\alpha$ fixed and $\beta_1$ varying. A model showing the behavior of pressure and density profiles for seeding radial pressure with theta component is presented. We can see from Fig. \ref{fig: press1} that the radial and transverse pressures decrease smoothly to the boundary at fixed $\alpha$. At increased $\beta_1$, the radial and tangential stresses decrease. When the coupling parameter increases, the radial and tangential pressures are suppressed. The disappearance of the radial pressure at a specific value of the radial coordinate defines the boundary of the stellar object. Near the centre of the fluid configuration, the radial and tangential pressures are similar, but the tangential pressure becomes dominant as the surface approaches. There is an interesting observation of switching tangential pressure within the core. As the coupling parameter increases, $p_t^{\text{eff}}$ decreases for some finite radii. Fig. \ref{fig: rho1} shows how the density profile behaves under symmetric teleparallel gravity for compact stars. The density in the central core region of a star behaves very similarly to what happens in the $\theta_0^0=\rho$ solution. Fig. \ref{fig: anisotropy1} illustrates the effective anisotropy of the $\theta^1_1=p_r$ solution. Using the left panel of the figure, we can see that the anisotropy parameter is positive throughout the stellar configuration, as seen in the figure. By increasing $\beta_1$, the force due to pressure anisotropy is strengthened, which forces the inwardly driven gravitational force to counteract the repulsive force. Interestingly, as $\beta_1$ increases, the anisotropy also increases, which strengthens the force due to pressure anisotropy.

    \begin{figure*}
        \centering
        \includegraphics[width=8cm,height=6.5cm]{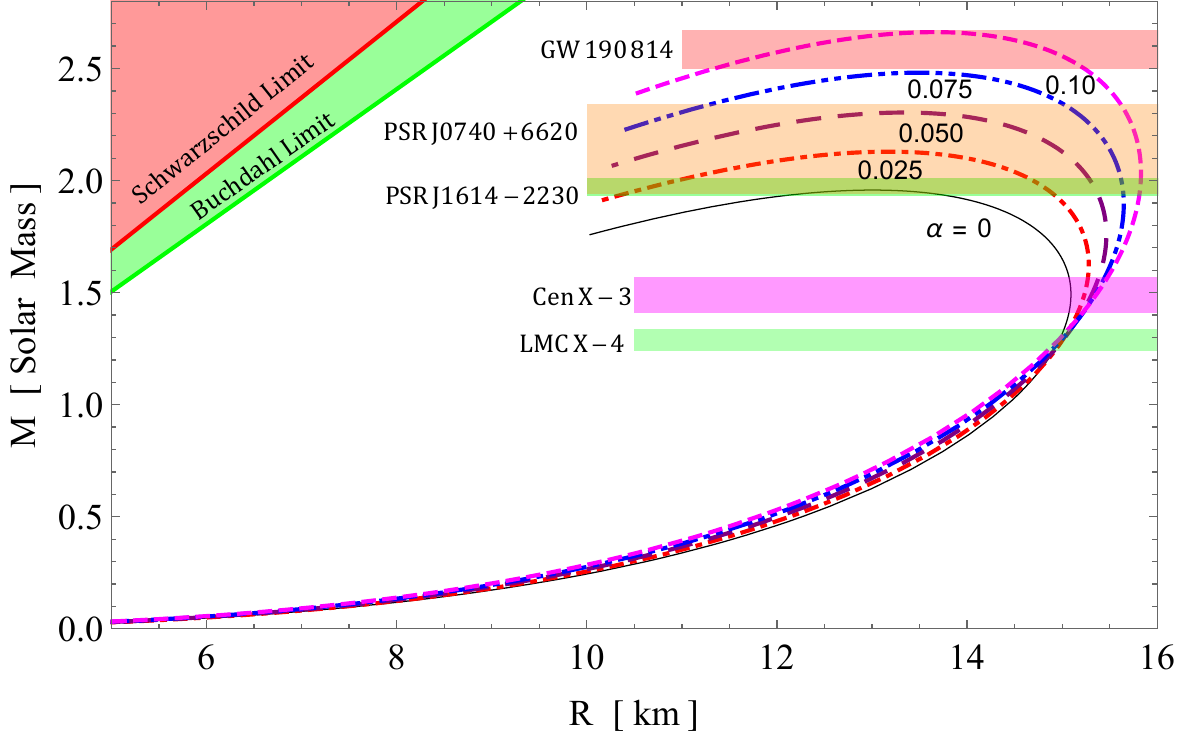}~~~ \includegraphics[width=8cm,height=6.5cm]{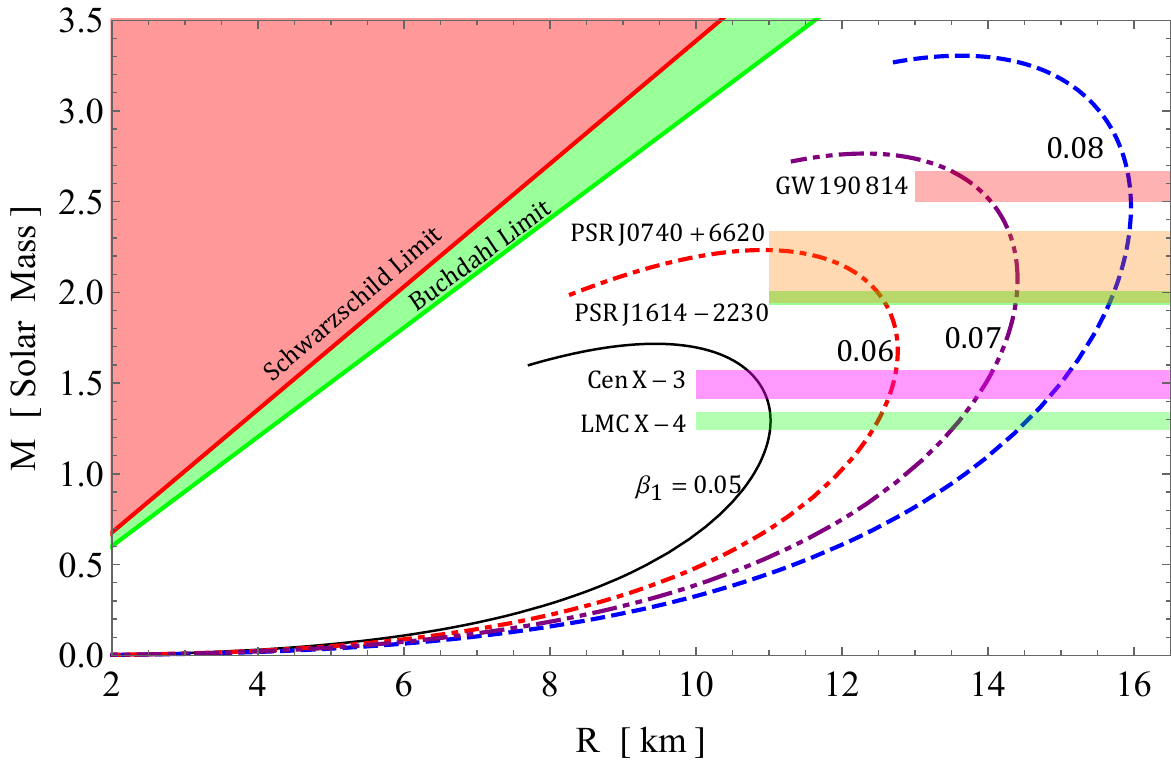}
        \caption{The radii measurement of observed compact objects via $M-R$ curves for different values of $\alpha$ and $\beta_1$  for solution \ref{solA} ($\rho=\theta^0_0$).} \label{m-r-1-a}
    \end{figure*}
    \begin{figure}
        \centering
        \includegraphics[width=8cm,height=6.5cm]{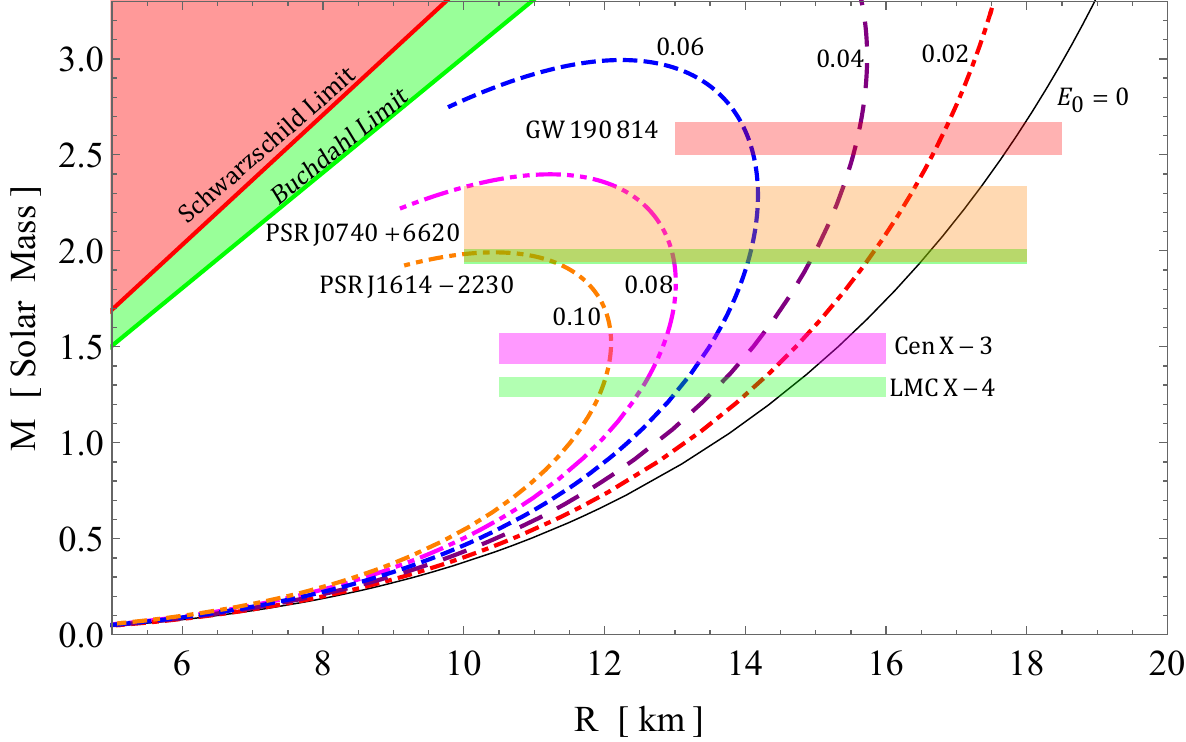}
        \caption{The radii measurement of observed compact objects via $M-R$ curves for different values of charge constant $E_0$ for solution \ref{solB} ($\rho=\theta^0_0$).} \label{m-r-1-E}
    \end{figure}
    
    \begin{figure*}
        \centering
        \includegraphics[width=8cm,height=6.5cm]{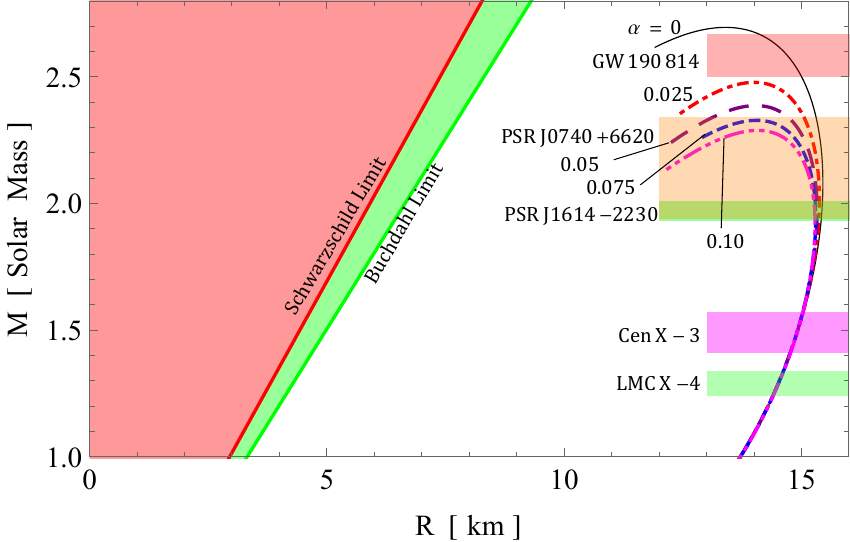}~~~\includegraphics[width=8cm,height=6.5cm]{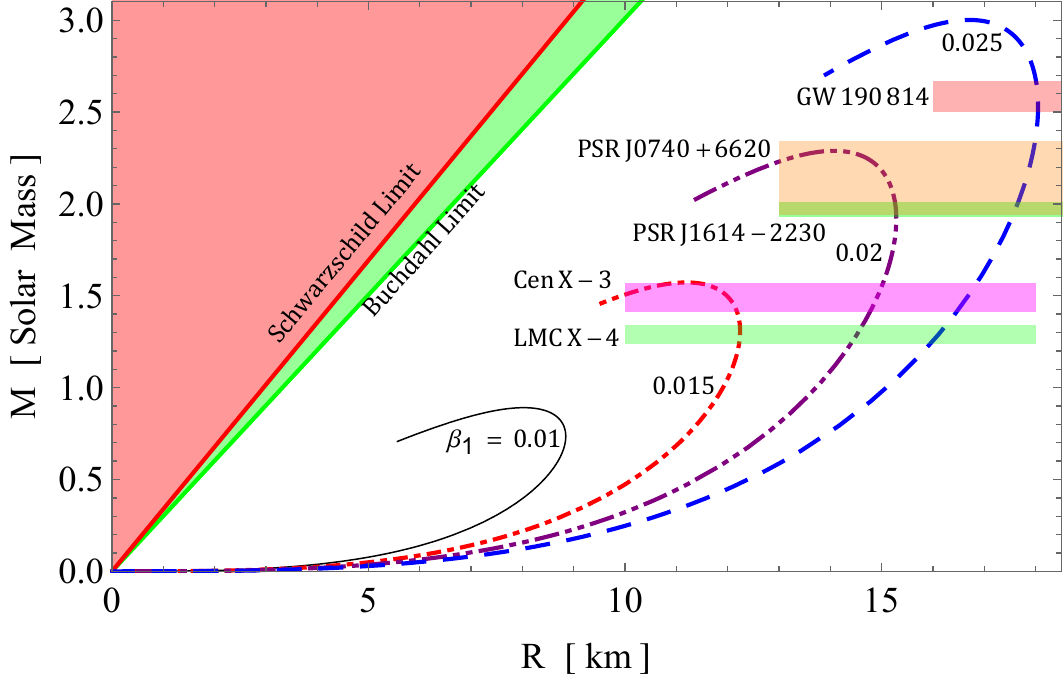}
        \caption{The radii measurement of observed compact objects via $M-R$ curves for different values of $\alpha$ and $\beta_1$  for solution \ref{solB} ($p_r=\theta^1_1$).} \label{m-r-2-a}
    \end{figure*}

    \begin{figure}
        \centering
        \includegraphics[width=8cm,height=6.5cm]{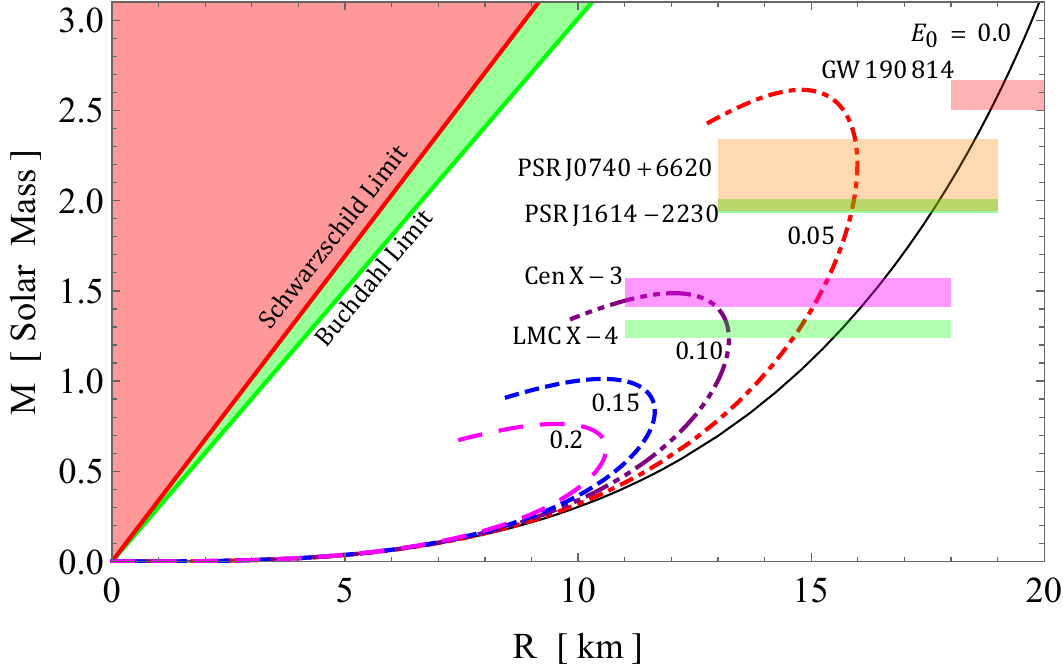}
        \caption{The radii measurement of observed compact objects via $M-R$ curves for different values of charge constant $E_0$ for solution \ref{solB} ($p_r=\theta^1_1$).} \label{m-r-2-E}
    \end{figure}

\section{Mass-radius constraints via \texorpdfstring{$M-R$}{} curves} \label{sec6}

In this section, we will focus on examining the effect of coupling parameters \texorpdfstring{$\alpha$ \text{and} $\beta_1$}{} as well as the electric charge parameter \texorpdfstring{$E_0$}{} in constraining the mass-radius relationship for different objects via \texorpdfstring{$M-R$}{} curves by exploiting both solutions \ref{solA} ($\theta_0^0 =\rho$ solution) and \ref{solB} ($\theta_1^1 =p_r$ solution).

\subsection{The effect of coupling parameters \texorpdfstring{$\alpha$ \text{and} $\beta_1$}{} on mass-radius relation for both solutions \texorpdfstring{\ref{solA} ($\theta_0^0 =\rho$) \text{and} \ref{solB} ($\theta_1^1 =p_r$)}{}}

Here, we first discuss the effect of coupling parameters \texorpdfstring{$\alpha$ \text{and} $\beta_1$} on constraining the mass-radius of four compact stars LMC X-4, Cen X-3, PSR J1614-2230, PSR J0740+6620 with masses ranging from  $1.29 ^{+ 0.05}_{-0.05}$\,M$_\odot$ to $2.14 ^{+ 0.2}_{-0.17}$\,M$_\odot$ along with the massive compact stars with masses in the range of 2.5 M$_\odot$ to 2.67 M$_\odot$ came from the GW190814 event for MIT-bag model using \texorpdfstring{$M-R$}{} diagrams for the $\theta_0^0 =\rho$ solution \ref{solA}. For drawing the total mass $M$ (normalized in M$_{\odot}$) vs the radius $R$ for both solutions ($\theta_0^0 =\rho$) and ($\theta_1^1 =p_r$), we select the values of parameters $\alpha$ and $\beta_1$. In this concern, we choose $\alpha\in[0.0,~0.10]$ and $\beta_1\in[0.05,~0.08]$ for the solution ($\theta_0^0 =\rho$, while  $\alpha\in~[0.0,~0.10]$ and $\beta_1\in[0.010,~0.025]$ for the other solution ($\theta_1^1 =p_r$). We have also included five horizontal strips of different colours, each representing the mass range of a distinct neutron star candidate for both solutions \ref{solA} ($\theta_0^0 =\rho$) and \ref{solB} ($\theta_1^1 =p_r$).

For the first solution \ref{solA} ($\theta_0^0 =\rho$), the Fig. \ref{m-r-1-a} shows that, with rising values of $\alpha-$(left panel) and $\beta_1-$(right panel), the maximum value of mass ($M_{max}$) gradually rises. The maximum allowable mass and corresponding radius for our first solution are shown in Table \ref{table1} for different values of $\alpha$ and $\beta_1$. It's noteworthy to notice that the maximum masses range from 1.95 M$_{\odot}$ to 2.65 M$_{\odot}$ for $\alpha$ and from 1.52 M$_{\odot}$ to 3.35 M$_{\odot}$ for $\beta_1$ while the radii corresponding to the maximum masses lie in 14.96 km to 15.64 km for $\alpha$ and in 10.71 km to 15.95 km for $\beta_1$. These considerations, together with the analysis above, lead us to the conclusion that higher $\beta_1$ generates more massive compact objects, whereas higher $\alpha$ generates compact objects that are less massive than those generated by $\beta_1$. This first solution ($\theta_0^0 =\rho$) makes the crucial discovery that for $\alpha=0.10$ and $\beta_1=0.07~\&~0.08$, we forecast the presence of a secondary component of the GW190814 event with mass $2.5-2.67$ M$_{\odot}$ lying in the hypothesized lower "\textit{mass gap}" in the context of $f(Q)$ gravity.

For the second solution \ref{solB} ($\theta_1^1 =p_r$), the Fig. \ref{m-r-2-a} shows that the maximum value of mass ($M_{max}$) progressively lowers for considering $\alpha-$(left panel) but gradually grows for considering $\beta_1-$(right panel) with growing values of $\alpha$ and $\beta_1$. For different values of $\alpha$ and $\beta_1$, the maximum allowable mass and corresponding radius for our second solution are displayed in Table \ref{table2}. It is interesting to note that while the radii for the maximum masses are 14.64 km to 15.50 km for $\alpha$ and in 12.01 km to 18.05 km for $\beta_1$, the maximum masses range from 2.3 M$_{\odot}$ to 2.7 M$_{\odot}$ and from 0.9 M$_{\odot}$ to 3.0 M$_{\odot}$ for $\alpha$ and $\beta_1$, respectively. These factors, together with the analysis above, lead us to the conclusion that higher $\beta_1$ generates objects that are less massive and more compact, whereas higher $\alpha$ generates objects that are less massive and less compact than those generated by $\beta_1$. The most interesting finding is that the presence of a secondary component of the GW190814 event with mass $2.5-2.67$ M$_{\odot}$ lying in the hypothesized lower "\textit{mass gap}" in the context of $f(Q)$ gravity has been achieved in the absence of deformation via decoupling constant $\alpha$ along with a slight deviation on constant parameter $\beta_1$.

In this sense, we would like to point out that the secondary mass of GW190814 is within the range of $2.5-2.67$ M$_{\odot}$, which is the lower "\textit{mass gap}" between known neutron stars and black holes \citep{Ozel:2012ax, Farr:2010tu, Ozel:2010su, Bailyn:1997xt}. It almost exceeds the mass of the $1.61-2.52$ M$_{\odot}$ primary component of GW190425, which is itself an outlier relative to the galactic population of binary neutron stars \citep{LIGOScientific:2020iuh}. It is heavier than the most massive pulsar in the Milky Way \citep{NANOGrav:2019jur}. Additionally, it is comparable to the millisecond pulsar PSR J1748-2021B \citep{Freire:2007jd}, whose mass is estimated to be $2.74 \pm 0.21$ M$_{\odot}$ with a 68\% confidence level.

\subsection{The effect of electric charge parameter \texorpdfstring{$E_0$}{} on mass-radius relation for both solutions \texorpdfstring{\ref{solA} ($\theta_0^0 =\rho$) \text{and} \ref{solB} ($\theta_1^1 =p_r$)}{}}

In addition to the above-mentioned effect of coupling parameters \texorpdfstring{$\alpha$ \text{and} $\beta_1$} on constraining the mass-radius of four compact stars LMC X-4, Cen X-3, PSR J1614-2230, and PSR J0740+6620 along with the massive compact stars which came from the GW190814 event for MIT-bag model using \texorpdfstring{$M-R$}{} diagrams for both solutions ($\theta_0^0 =\rho$) and ($\theta_1^1 =p_r$), we discuss here another important physical feature on the stellar systems which is the effect of electric charge parameter \texorpdfstring{$E_0$}{} on mass-radius relation for the same solutions \ref{solA} ($\theta_0^0 =\rho$) and \ref{solB} ($\theta_1^1 =p_r$). Technically the effect of electric charge parameter \texorpdfstring{$E_0$}{} on constraining the mass-radius relation for both solutions $\theta_0^0 =\rho$ and $\theta_1^1 =p_r$ can be described by a slight change in its value specifically from $E_0=0.00$ to $0.10$ and $E_0=0.00$ to $0.20$ for $\theta_0^0 =\rho$ and $\theta_1^1 =p_r$, respectively. According to Figs. \ref{m-r-1-E} and \ref{m-r-2-E}, the maximum value of mass ($M_{max}$) gradually drops with increasing values of $E_0$ for both solutions. Table \ref{table3} for different values of $E_0$ displays the maximum allowable mass and related radius for $\theta_0^0 =\rho$ and $\theta_1^1 =p_r$, respectively. It is captivating to point out that the maximum masses range from 1.95 M$_{\odot}$ to 3.00 M$_{\odot}$ with $E_0~\in~[0.06,~0.10]$ for $\theta_0^0 =\rho$ and from 0.75 M$_{\odot}$ to 2.60 M$_{\odot}$ with $E_0~\in~[0.06,~0.10]$ for $\theta_1^1 =p_r$ while the radii corresponding to the maximum masses lie in 10.55 km to 14.18 km with $E_0~\in~[0.6.0,~0.10]$ for $\theta_0^0 =\rho$ and 9.75 km to 15.98 km with $E_0~\in~[0.06,~0.10]$ for $\theta_1^1 =p_r$. These factors lead us to the conclusion that higher $E_0$ generates less massive compact objects, whereas lower $E_0$ generates compact objects that are more massive for both solutions according to the involved values of $E_0$. The masses and radii, however, do not reach their maximum values in $E_0<0.06$ for $\theta_0^0 =\rho$ or $E_0<0.05$ for $\theta_1^1 =p_r$, respectively.

The $f(Q)$ gravitational theory has been instrumental in describing one of the components of the event GW190814 \cite{LIGOScientific:2020zkf}, which displayed the coalescence of a black hole with a mass of $22.20-24.30$ M$_{\odot}$ and a compact object with a mass of $2.50-2.67$ M$_{\odot}$, the latter of which may have been an exceptionally high-mass neutron star or a low-mass black hole. 

\begin{table*}
    \centering
    \caption{The predicted radii of compact stars LMC X-4, Cen X-3, PSR J1614-2230, PSR J0740+6620, and GW190814 for MIT-bag model (see Fig.\ref{m-r-1-a}). }\label{table1}
    \scalebox{0.8}
    {\begin{tabular}{| *{11}{c|} }
\hline
    {Objects} & {$\frac{M}{\text{M}_\odot}$}   & \multicolumn{9}{c|}{{Predicted $R$ km}}   \\
    \cline{3-11}
    && \multicolumn{9}{c|}{ For solution $\rho=\theta^0_0$}  \\
    \cline{3-11}
    &  &  $\alpha=0$ & $\alpha=0.025$  & $\alpha=0.05$  & $\alpha=0.075$  &$\alpha=0.10$  & $\beta_1=0.05$ & $\beta_1=0.06$ & $\beta_1=0.07$  & $\beta_1=0.08$   \\ \hline
    LMC X-4 \citep{star3}  &  1.29 $\pm$ 0.05  & $14.97_{-0.01}^{+0.01}$  &  $14.98_{-0.01}^{+0.01}$  &    $15.05_{-0.01}^{+0.01}$  &  $15.04_{-0.01}^{+0.01}$  &  $14.79_{-0.01}^{+0.01}$ & $11.01_{-0.01}^{+0.01}$ & $12.49_{-0.08}^{+0.06}$ & $13.61_{-0.11}^{+0.09}$ & $14.53_{-0.11}^{+0.11}$   \\
\hline
    Cen X-3 \citep{star3} & 1.49$\pm$0.08  &  $15.10_{-0.03}^{+0.01}$ & 15.23$_{-0.07}^{+0.04}$ & 15.32$_{-0.12}^{+0.06}$ & 15.34$_{-0.12}^{+0.11}$ & 15.36$_{-0.15}^{+0.12}$ &  $10.88_{-0.17}^{+0.08}$ & 12.70$_{-0.08}^{+0.03}$ & 13.98$_{-0.16}^{+0.07} $ &  $14.98_{-0.17}^{+0.03}$ \\
\hline
    PSR J1614-2230 \citep{star1} & 1.97$\pm$0.04  & -  & 14.82$_{-0.19}^{+0.11}$ & 15.33$_{-0.05}^{+0.05}$ & 15.64$_{-0.01}^{+0.01}$ & 15.82$_{-0.01}^{+0.01}$ & - &  $12.53_{-0.05}^{+0.07}$ & 14.37$_{-0.01}^{+0.01}$ & 15.69$_{-0.03}^{+0.05} $ \\
\hline
    PSR J0740+6620 \citep{Cromartie} & $2.14^{+0.2}_{-0.17}$ & - & -  & 14.95$_{-0.35}^{+0.27}$ & 15.49$_{-0.47}^{+0.15}$ & 15.80$_{-0.18}^{+0.02}$ & - & 12.01$_{-0.0}^{+0.57}$ & 14.38$_{-0.08}^{+0.01}$ &  $15.84_{-0.19}^{+0.09}$  \\
\hline
    GW190814 \citep{wenbin} & 2.5-2.67 & - & -  & - & - & 14.90$_{-1.2}^{+0.33}$ & - & - & 13.86$_{-0.33}^{+0.14}$ & 15.94$_{-0.03}^{+0.01}$   \\
\hline
    \end{tabular}}
\end{table*}

\begin{table*}
    \centering
    \caption{The predicted radii of compact stars LMC X-4, Cen X-3, PSR J1614-2230, PSR J0740+6620, and GW190814 for MIT-bag model (see Fig.\ref{m-r-2-a}). }\label{table2}
    \scalebox{0.8}
    {\begin{tabular}{| *{11}{c|} }
\hline
    {Objects} & {$\frac{M}{\text{M}_\odot}$}   & \multicolumn{9}{c|}{{Predicted $R$ km}}   \\
    \cline{3-11}
    && \multicolumn{9}{c|}{ For solution $p_r=\theta^1_1$}  \\
    \cline{3-11}
    &  &  $\alpha=0$ & $\alpha=0.025$  & $\alpha=0.05$  & $\alpha=0.075$  &$\alpha=0.10$  & $\beta_1=0.01$ & $\beta_1=0.015$ & $\beta_1=0.02$ & $\beta_1=0.025$   \\
\hline
    LMC X-4 \citep{star3}  &  1.29 $\pm$ 0.05  & $14.52_{-0.10}^{+0.12}$  &  $14.52_{-0.10}^{+0.12}$  &    $14.52_{-0.10}^{+0.12}$  &  $14.52_{-0.10}^{+0.12}$  &  $14.52_{-0.10}^{+0.12}$ & - & $12.22_{-0.01}^{+0.02}$ & $14.52_{-0.09}^{+0.14}$ & $16.11_{-0.06}^{+0.12}$   \\
\hline
    Cen X-3 \citep{star3} & 1.49$\pm$0.08  &  $14.96_{-0.19}^{+0.08}$ & 14.95$_{-0.19}^{+0.08}$ & 14.94$_{-0.19}^{+0.08}$ & 14.93$_{-0.11}^{+0.08}$ & 14.92$_{-0.19}^{+0.08}$ &  - & 11.99$_{-0.58}^{+0.20}$ & 14.93$_{-0.15}^{+0.10} $ &  $16.66_{-0.19}^{+0.22}$ \\
\hline
    PSR J1614-2230 \citep{star1} & 1.97$\pm$0.04  & 15.44$_{-0.02}^{+0.01}$  & 15.38$_{-0.01}^{+0.01}$ & 15.33$_{-0.01}^{+0.01}$ & 15.31$_{-0.01}^{+0.01}$ & 15.28$_{-0.01}^{+0.01}$ & - &  - & 15.29$_{-0.02}^{+0.01}$ & 17.62$_{-0.03}^{+0.03} $ \\
\hline
    PSR J0740+6620 \citep{Cromartie} & $2.14^{+0.2}_{-0.17}$ & 15.45$_{-0.02}^{+0.05}$ & 15.34$_{-0.28}^{+0.04}$  & 15.24$_{-0.48}^{+0.09}$ & 15.16$_{-0.0}^{+0.14}$ & 15.11$_{-0.0}^{+0.17}$ & - & - & 15.12$_{-0.0}^{+0.17}$ &  $17.84_{-0.27}^{+0.14}$  \\
\hline
    GW190814 \citep{wenbin} & 2.5-2.67 & - & -  & - & - & 14.76$_{-0.65}^{+0.24}$ & - & - & - & 18.04$_{-0.03}^{+0.01}$   \\
\hline
    \end{tabular}}
\end{table*}


\begin{table*}
    \centering
    \caption{The predicted radii of compact stars LMC X-4, Cen X-3, PSR J1614-2230, PSR J0740+6620, and GW190814 for MIT-bag model (see Figs. \ref{m-r-1-E}, \ref{m-r-2-E}). }\label{table3}
    \scalebox{0.8}
    {\begin{tabular}{| *{11}{c|} }
\hline
    {Objects} & {$\frac{M}{\text{M}_\odot}$}   & \multicolumn{9}{c|}{{Predicted $R$ km}}   \\
    \cline{3-11}
    & & \multicolumn{6}{c|}{ For solution $\rho=\theta^0_0$} &\multicolumn{3}{c|}{ For solution $p_r=\theta^1_1$}   \\
    \cline{3-11}
    &  &  $E_0=0$ & $E_0=0.02$  & $E_0=0.04$  & $E_0=0.06$  & $E_0=0.08$  & $E_0=0.1$ & $E_0=0$ & $E_0=0.05$  & $E_0=0.1$\\ 
\hline
    LMC X-4 \citep{star3}  &  1.29 $\pm$ 0.05  & $14.65_{-0.16}^{+0.18}$  &  $14.12_{-0.18}^{+0.13}$  &    $13.59_{-0.14}^{+0.12}$  &  $13.08_{-0.12}^{+0.10}$  &  $12.54_{-0.06}^{+0.08}$ & $11.98_{-0.06}^{+0.04}$ & $15.65_{-0.17}^{+0.17}$ & $14.76_{-0.15}^{+0.07}$ & $13.21_{-0.04}^{+0.02}$   \\
\hline
    Cen X-3 \citep{star3} & 1.49$\pm$0.08  &  $15.29_{-0.25}^{+0.23}$ & 14.67$_{-0.18}^{+0.21}$ & 14.09$_{-0.20}^{+0.16}$ & 13.46$_{-0.15}^{+0.14}$ & 12.82$_{-0.11}^{+0.07}$ &  $12.11_{-0.04}^{+0.01}$ & 16.32$_{-0.24}^{+0.23}$ & 15.18$_{-0.12}^{+0.19} $ &  - \\
\hline
    PSR J1614-2230 \citep{star1} & 1.97$\pm$0.04  & 16.57$_{-0.07}^{+0.07}$  & 15.77$_{-0.04}^{+0.06}$ & 14.98$_{-0.07}^{+0.04}$ & 14.04$_{-0.01}^{+0.04}$ & 12.98$_{-0.02}^{+0.01}$ & $10.96_{-0.41}^{+0.32}$ &  $17.66_{-0.05}^{+0.66}$ & 15.91$_{-0.05}^{+0.01}$ & -  \\
\hline
    PSR J0740+6620 \citep{Cromartie} & $2.14^{+0.2}_{-0.17}$ & 16.94$_{-0.46}^{+0.43}$ & 16.10$_{-0.37}^{+0.33}$  & 15.18$_{-0.27}^{+0.22}$ & 14.17$_{-0.13}^{+0.01}$ & 12.77$_{-0.22}^{+0.36}$ & - & $18.09_{-0.50}^{+0.43}$ & $15.97_{-0.09}^{+0.01}$ & -   \\
\hline
    GW190814 \citep{wenbin} & 2.5-2.67 & $17.79_{-0.12}^{+0.20}$ & $16.76_{-0.09}^{+0.13}$  & $15.60_{-0.07}^{+0.03}$ & $14.09_{-0.16}^{+0.03}$ & - & - & $18.99_{-0.15}^{+0.16}$ & $15.31_{-0.0}^{+0.40}$ & -   \\
\hline
    \end{tabular}}
\end{table*}

\section{The transfer of energy between fluid systems for \texorpdfstring{$T_{\lowercase{ij}}$}{} and \texorpdfstring{$\theta_{\lowercase{ij}}$}{}} \label{sec7}

Let us now talk about the need for energy exchange in relation to extended gravitational decoupling. In his work, Ovalle (\cite{Ovalle_2019_788}) demonstrated that the successful decoupling of the sources $T_{ij}$ and $\theta_{ij}$ depends on the presence of energy exchange between them. The equations of motion of the line element in $f(Q)$-gravity, denoted by the symbol $\mathcal{G}_{ij}$, can be thought of as follows: 
    \begin{eqnarray}
        && \hspace{-1.3cm} \mathcal{G}_{ij}=  \frac{2}{\sqrt{-g}}\nabla_k\left(\sqrt{-g}\,f_Q\,P^k_{\,\,\,\,i j}\right)+\frac{1}{2}g_{i j}f 
        +f_Q\big(P_{i\,k l}\,Q_j^{\,\,\,\,k l}\nonumber\\
        && \hspace{-0.5cm} -2\,Q_{k l i}\,P^{k l}_{\,\,\,\,\,j}\big) = - (T^{\text{eff}}_{ij}+E_{ij}) = -(T_{ij}+\alpha ~\theta_{ij}+E_{ij}).~~~\label{eq94}
    \end{eqnarray}

The conservation equation can be found by Bianchi identity $ \bigtriangledown^i\mathcal{G}_{ij}=0$, given by
    \begin{small}
    \begin{eqnarray}
        && \hspace{-2cm}-\frac{H^{\prime}}{2}\,(T^0_{0}-T^1_{1})+(T^1_{1})^{\prime}-\frac{2}{r}\,(T^2_{2}-T^1_{1})-\frac{\alpha \xi^{\prime}}{2} (T^0_{0}-T^1_{1}) \nonumber\\&&\hspace{-2cm}-\frac{H^{\prime}}{2}\,(E^0_{0}-E^1_{1})+(E^1_{1})^{\prime}-\frac{2}{r}\,(E^2_{2}-E^1_{1})-\frac{\alpha \xi^{\prime}}{2} (E^0_{0}-E^1_{1}) \nonumber\\- 
        && \hspace{-2cm} \frac{\alpha \Phi'}{2} \Big(\theta^0_0-\theta^1_1\Big) +\alpha \Big(\theta^{1}_{1}\Big)^{\prime}-\frac{2 \alpha}{r}\,\Big(\theta^{2}_{2}-\theta^{1}_{1}\Big)=0.~~~~~~\label{eq.76}
    \end{eqnarray} 
    \end{small}
A noteworthy aspect in the context of the TOV equation for $f(\mathcal{Q})$ for the static and spherically symmetric line element is the same as the general relativity context under the linear functional form.  This guarantees that the metric (\ref{eq.45}) should meet the relevant Bianchi identity for $\mathcal{G}^{\{H,W\}}_{ij}$. Again, this implies that the space-time geometry $H, W$ of equation (\ref{eq.44}) should preserve the energy-momentum tensor $T_{ij}$. Consequently, one can offer,     
    \begin{eqnarray}
        && \hspace{-6cm} \bigtriangledown^{\{H,W\}}_i \, (T^{i}_ {j}+E^{i}_ {j})=0. \label{eq.77} 
    \end{eqnarray}

It is also instructive, in connection to equation (\ref{eq.11}), that
    \begin{small}
    \begin{eqnarray}
        \bigtriangledown_i \, T^{i}_{ j} =  \bigtriangledown^{\{H,W\}}_i \, (T^{i}_{j}+E^i_j)-\frac{\alpha \,\xi'}{2} (T^0_0-{T}^1_1) \delta^1_j-\frac{\alpha \,\xi'}{2} (E^0_0-{E}^1_1) \delta^1_j. \label{eq.78}
    \end{eqnarray}
    \end{small}
As a linear combination of the Einstein field equations (\ref{eq.41})--(\ref{eq.43}), we obtain from equation (\ref{eq.77}) the following explicit form:
    \begin{eqnarray}
        && \hspace{-0.8cm}-\frac{H^{\prime}}{2}\,(T^0_{0}-T^1_{1})+(T^1_{1})^{\prime}-\frac{2}{r}\,(T^2_{2}-T^1_{1})-\frac{H^{\prime}}{2}\,(E^0_{0}-E^1_{1})+(E^1_{1})^{\prime} \nonumber\\&&-\frac{2}{r}\,(E^2_{2}-E^1_{1})=0.~~~~\label{eq.79}
    \end{eqnarray}
This immediately shows that the sources $T_{ij}$ and $E_{ij}$ may be well-defined decoupled from the system of equations (\ref{eq.41})--(\ref{eq.43}), and eventually, based on the condition (\ref{eq.77}), one can derive the following forms from equation (\ref{eq.76}): 
    \begin{eqnarray}
        && \hspace{-1.7cm} \bigtriangledown_i \, T^{i}_{j} =-\frac{\alpha\, \xi'}{2} (T^0_0-T^1_1) \delta^1_j, \label{eq.80}\\
        && \hspace{-1.7cm} \bigtriangledown_i \, E^{i}_{j} =-\frac{\alpha\, \xi'}{2} (E^0_0-E^1_1) \delta^1_j =0,~~ \text{as}~~\big[E^0_0=E^1_1=\frac{q^2}{r^4}\big].~~~~ 
    \end{eqnarray}
and 
    \begin{eqnarray}
        && \hspace{-5.8cm} \bigtriangledown_i \, \theta^{i}_{ j}= \frac{\alpha\, \xi'}{2} (T^0_0-T^1_1) \delta^1_j. \label{eq.82}
    \end{eqnarray}
At this point, we would to mentioned that decoupling with the electric field and new sector does not demand any restriction, while decoupling can be accomplished as long as there is an energy transfer between the sources $T_{ij}$ and $\theta_{ij}$. In this regard, the energy exchange between the sources can be described as follows using the works \citep{Ener1, Contreras_2022_82},  
    \begin{eqnarray}
        && \hspace{-6.8cm} \Delta \mathcal{E}=\frac{\xi^\prime }{2} \big(p_r + \rho\big). \label{eq.83}
    \end{eqnarray}
Since $p_r$ and $\rho$ are two positive physical quantities, the above equation (\ref{eq.83}) can help to examine the following situations: (i) if $\xi^\prime>0$, then $\Delta \mathcal{E}>0$, which means $\bigtriangledown_i \, \theta^{ i}_{j}>0$, i.e. the new source $ \theta_{i j}$ supplies energy to the surroundings, and (ii) if $\xi^\prime<0$, then $\Delta \mathcal{E}<0$, which implies $\bigtriangledown_i \, \theta^ {i}_{ j}<0$, i.e. the new source $ \theta_{i j}$ draws energy from the environment.

It should be noted that the temporal component is the same for both solutions; therefore, the expressions of the energy exchange are the same in both cases, but the amount of energy exchange is different in both cases. If we now use the expressions for seed pressure and seed density together with the time deformation function $\xi$, we find:

    \begin{figure*} 
        \centering 
        \includegraphics[width=4.9cm,height=4.8cm]{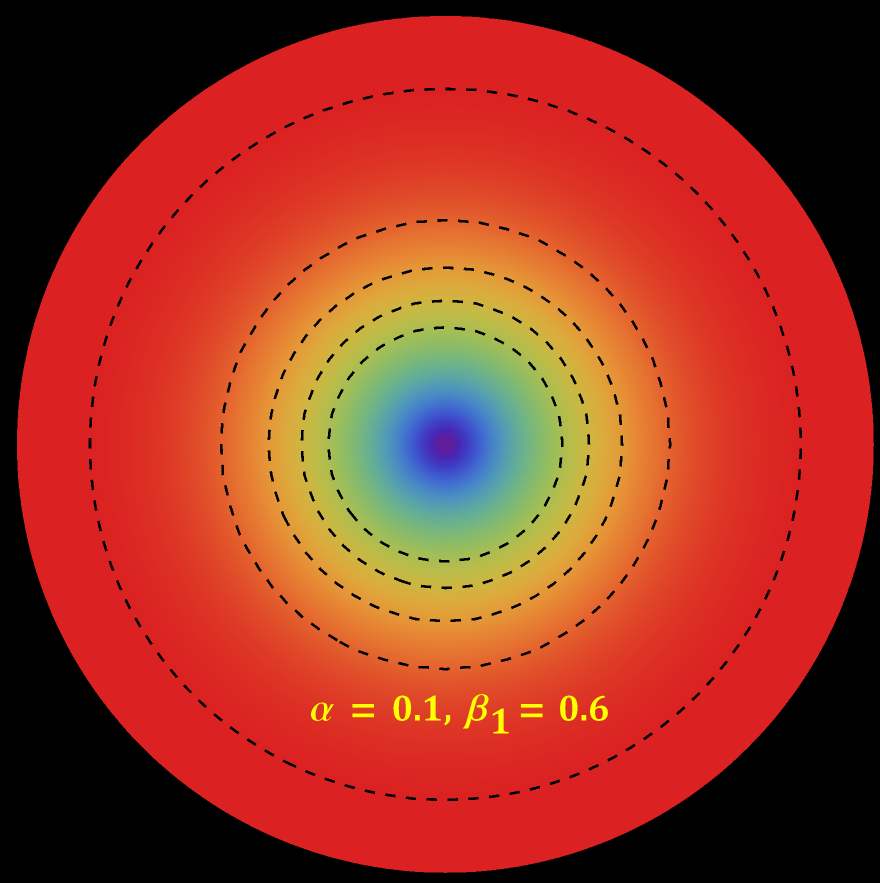} \includegraphics[width=0.9cm,height=4.6cm]{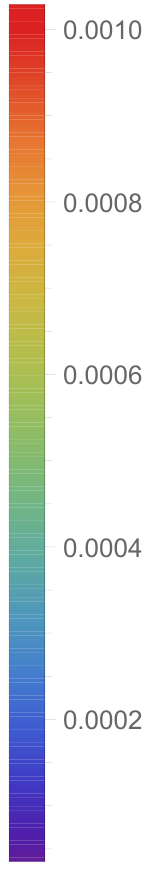} \,\includegraphics[width=4.8cm,height=4.6cm]{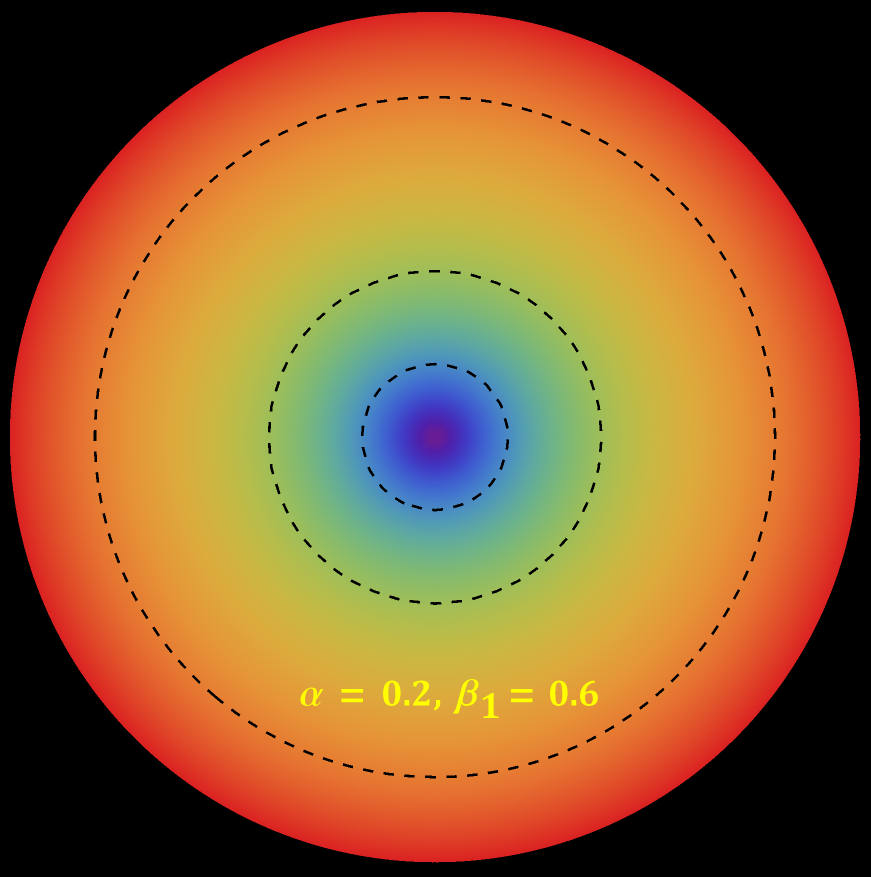} \includegraphics[width=0.9cm,height=4.6cm]{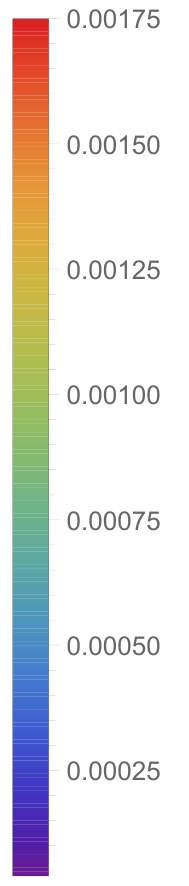}\, \includegraphics[width=4.8cm,height=4.6cm]{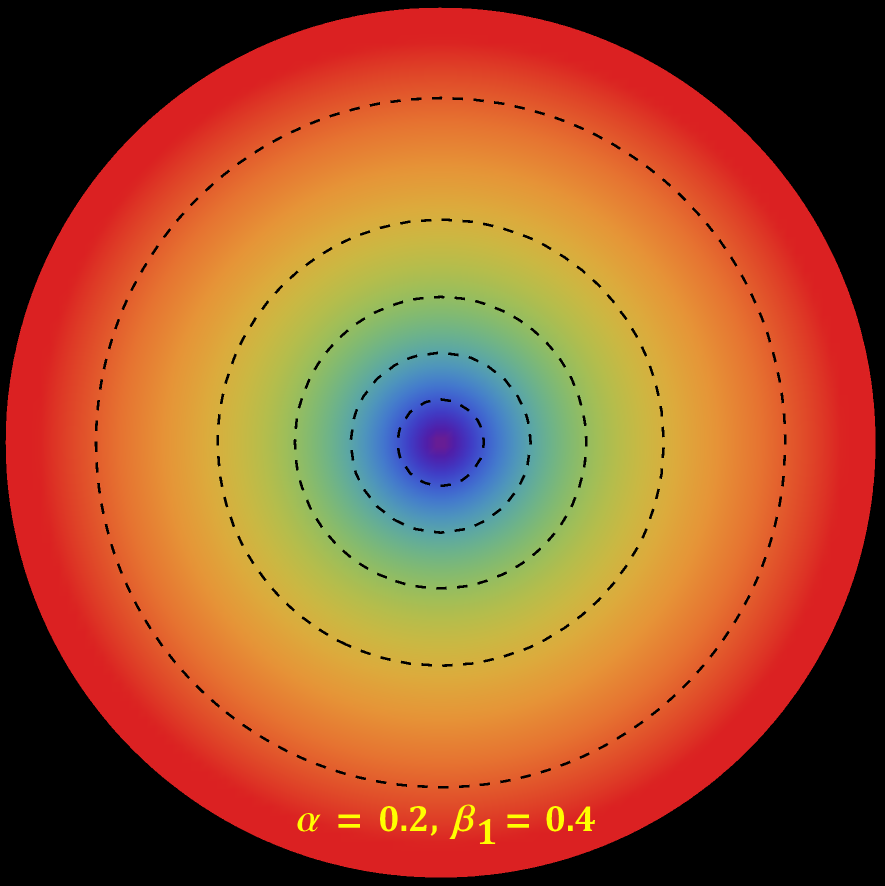} \includegraphics[width=0.9cm,height=4.5cm]{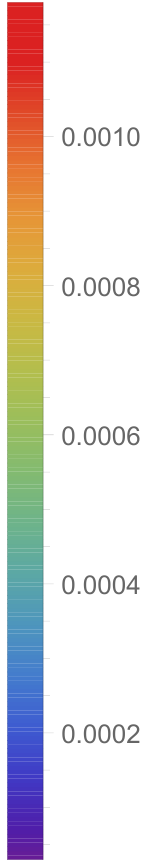}
        \caption{The flow of $energy-exchange$ ($\Delta\mathcal{E}$ - in $\text{km}^{-3}$) between the fluid distributions for the solution \ref{solA} ($\rho=\theta^0_0$)}
        \label{Ener-1}
    \end{figure*}

    \begin{figure*} 
        \centering 
         \includegraphics[width=5cm,height=4.9cm]{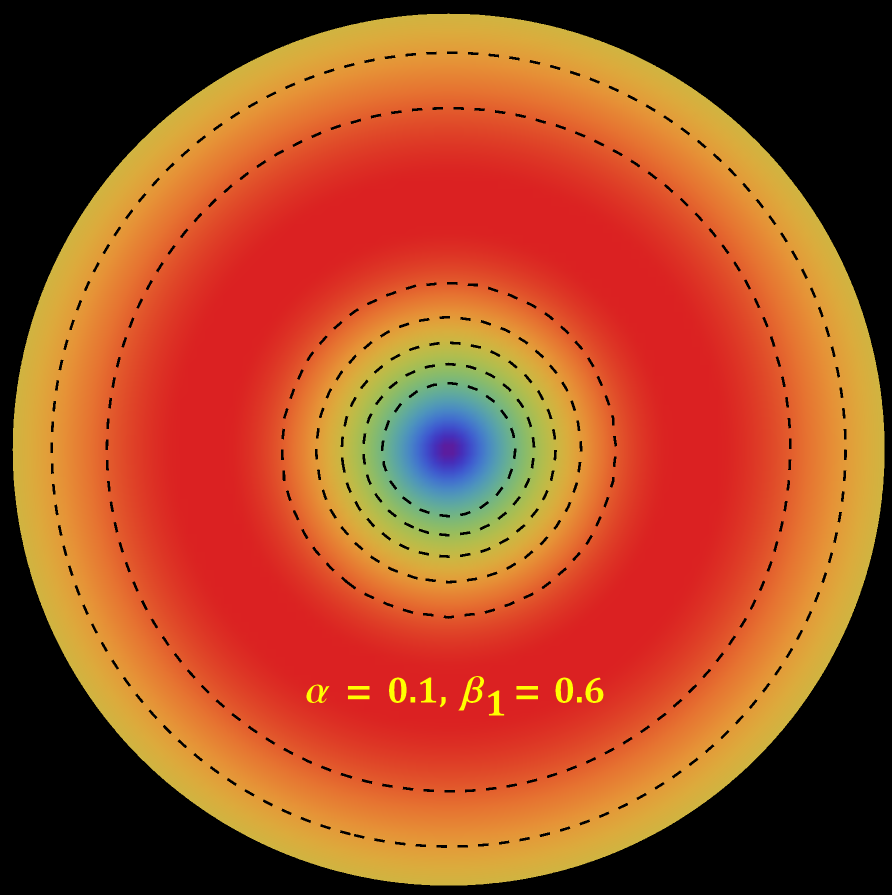} \includegraphics[width=1cm,height=4.7cm]{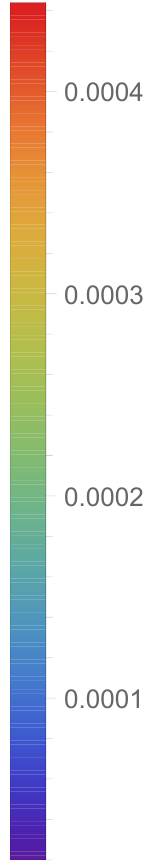} ~~~~~\includegraphics[width=5cm,height=4.9cm]{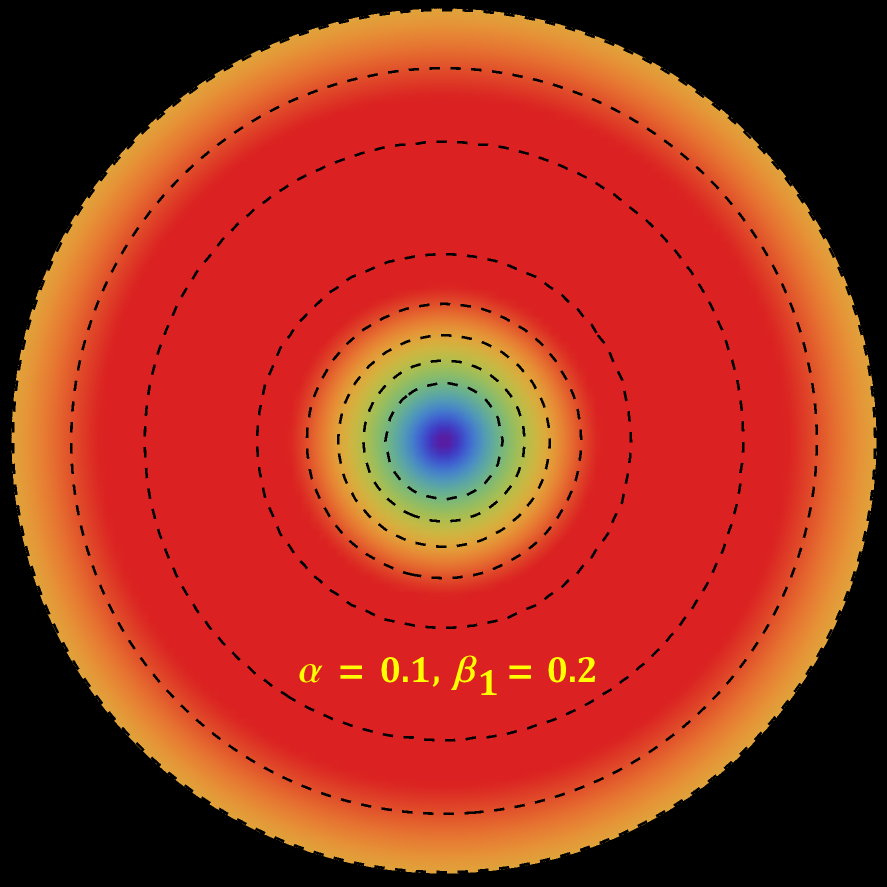} \includegraphics[width=1cm,height=4.7cm]{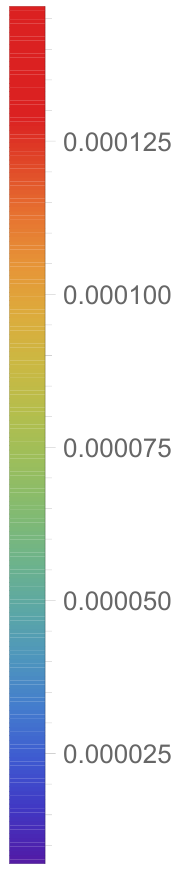}
        \caption{The flow of $energy-exchange$ ($\Delta\mathcal{E}$ - in $\text{km}^{-3}$) between the fluid distributions for the solution \ref{solB} ($p_r=\theta^1_1$).}
        \label{Ener-2}
    \end{figure*}

\textbf{For Solution \ref{solA}: ($\rho=\theta^0_0$)}

    \begin{eqnarray}
        &&\hspace{-0.5cm} \Delta \mathcal{E}= \frac{1}{9 \beta_1 K \left(C r^2+1\right)^3 \left(C r^2+K\right)}\Big[2 r \Big(2 \mathcal{B}_g K \left(C r^2+1\right)^2+C^2 r^2 \nonumber\\
        &&\hspace{-0.1cm} \left(2 \beta_1+2 E_0 K-2 \beta_1 K+\beta_2 K r^2\right)+C \left(6 \beta_1-6 \beta_1 K+2 \beta_2 K r^2\right) \nonumber\\
        &&\hspace{-0.5cm} +\beta_2 K\Big) \Big(2 \mathcal{B}_g K \left(C r^2+1\right)^2+C^2 r^2 \big(2 \beta_1+2 E_0 K-2 \beta_1 K+\beta_2 K r^2\big) \nonumber\\
        &&\hspace{0.2cm} +C \left(3 \beta_1-3 \beta_1 K+2 \beta_2 K r^2\right)+\beta_2 K\Big)\Big]\label{eq103}
    \end{eqnarray}

where, 
    \begin{small}
    \begin{eqnarray}
        &&\hspace{-0.5cm} \mathcal{B}_g=-\Big[C^3 R^4 \Big(-12 \alpha  (\alpha +1) E_0^2 K^2+E_0 K \big(-3 \beta_1 \big(4 \alpha ^2 (K-1) \nonumber\\
        &&\hspace{0.0cm} +\alpha  (4 K-15)-2\big)-2 \alpha  (\alpha +1) \beta_2 K R^2\big)+2 \alpha  (\alpha +1) \beta_2^2 K^2 R^4 \nonumber\\
        &&\hspace{0.0cm} +6 (\alpha +1) \beta_1^2 (K-1) (4 \alpha  (K-1)-1)+\beta_1 \beta_2 K R^2 [-16 \alpha ^2 \nonumber\\
        &&\hspace{0.0cm} \times (K-1)+\alpha  (25-16 K)+3]\Big)+C^2 R^2 \Big(E_0 K \big(-9 \alpha  (2 \alpha +1) \beta_1 \nonumber\\
        &&\hspace{0.0cm} +3 \left(6 \alpha ^2+17 \alpha +2\right) \beta_1 K-8 \alpha  (\alpha +1) \beta_2 K R^2\big)+6 \alpha  (\alpha +1) \beta_2^2  \nonumber\\
        &&\hspace{-0.1cm} K^2 R^4+6 (\alpha +1) \beta_1^2 (K-1) (-6 \alpha +(6 \alpha -1) K-3)+\beta_1 \beta_2 K R^2  \nonumber\\
        &&\hspace{-0.2cm} \Big(34 \alpha ^2+52 \alpha +\left(3-34 \alpha ^2-25 \alpha\right) K+6\Big)\Big) -3 C K \Big(\alpha  E_0 K \big(2 (\alpha +1)  \nonumber\\
        &&\hspace{0.0cm} \beta_2 R^2- 3 \beta_1\big)+6 (\alpha +1) \beta_1^2 (K-1)-2 \alpha  (\alpha +1) \beta_2^2 K R^4+\beta_1 \beta_2 R^2 \nonumber\\
        &&\hspace{0.2cm} \times \left(-6 \alpha ^2-9 \alpha +6 \alpha ^2 K-2 K-1\right)\Big)+\beta_2 K^2  \big((9 \alpha +3) \beta_1+2 \alpha  \nonumber\\
        &&\hspace{0.0cm} \times (\alpha +1) \beta_2 R^2\big)\Big]\Big[4 (\alpha +1) K \left(C R^2+1\right)^2 \Big(C R^2 \big(-3 \alpha  E_0 K+\beta_1 \nonumber\\
        &&\hspace{0.0cm} \times (6 \alpha -6 \alpha  K+6)+\alpha  \beta_2 K R^2\big)+K \left(6 \beta_1+\alpha  \beta_2 R^2\right)\Big)\Big]^{-1}, \nonumber
    \end{eqnarray}
    \end{small}

\textbf{For Solution \ref{solB}: ($p_r=\theta^1_1$)}

    \begin{eqnarray}
        &&\hspace{-0.7cm} \Delta \mathcal{E}= \frac{1}{9 \beta_1 K \left(C r^2+1\right)^3 \left(C r^2+K\right)}\Big[2 r \Big(2 \mathcal{B}_g K \left(C r^2+1\right)^2+C^2 r^2 \nonumber\\
        &&\hspace{0.1cm} \times \left(2 \beta_1+2 E_0 K-2 \beta_1 K+\beta_2 K r^2\right)+C \big(6 \beta_1-6 \beta_1 K+2 \beta_2 \nonumber\\
        &&\hspace{0.1cm}  K r^2\big)+\beta_2 K\Big) \Big(2 \mathcal{B}_g K \left(C r^2+1\right)^2+C^2 r^2 \big(2 \beta_1+2 E_0 K- \nonumber\\
        &&\hspace{0.1cm} 2 \beta_1 K+\beta_2 K r^2\big)+C \left(3 \beta_1-3 \beta_1 K+2 \beta_2 K r^2\right)+\beta_2 K\Big)\Big], 
    \end{eqnarray} 

where, 
    \begin{eqnarray}
        &&\hspace{-1.4cm} \mathcal{B}_g=\frac{-3}{4} \Bigg[\frac{C^2 R^2 \left(2 E_0 K-2 \beta_1 (K-1)+\beta_2 K R^2\right)}{6 K \left(C R^2+1\right)^2}- \nonumber\\
        &&\hspace{0.1cm} \frac{\beta_1 C (K-1)}{K \left(C R^2+1\right)^2}+\frac{\beta_2 C R^2}{3 \left(C R^2+1\right)^2}+\frac{\beta_2}{6 \left(C R^2+1\right)^2}\Bigg]. \nonumber
    \end{eqnarray}

The energy exchange between fluid distributions for both solutions \ref{solA} and \ref{solB}. The Fig. \ref{Ener-1} is plotted for $\alpha=0.1$ and $\beta_1=0.6$ (left panel), $\alpha=0.2$ and $\beta_1=0.6$ (middle panel), and $\alpha=0.2$ and $\beta_1=0.4$ (right panel) corresponding to solution \ref{solA}. In all cases, $\Delta \mathcal{E}$ is positive and increasing from centre to boundary. Therefore, the new source supplies energy to the environment, and the maximum value of $\Delta E$ is achieved at the boundary in all cases. It is noted that when the decoupling constant $\alpha$ increases, the value of $\Delta \mathcal{E}$ also increases. On the other hand, when we compare the last two panels of the figure, it is found that $\Delta \mathcal{E}$ is decreasing when constant $\beta_1$ decreases. Furthermore, Fig. \ref{Ener-2} shows the amount of energy exchange between new and seed sources for solution \ref{solB}. It is observed that when $\beta_1$ decreases the $\Delta \mathcal{E}$, while there is no impact of the decoupling constant $\alpha$ on the $\Delta \mathcal{E}$ for the second solution. Also, one can note that $\Delta \mathcal{E}$ increases from centre to boundary, but its maximum value exists inside the star rather than on the boundary. 

Finally, we conclude that both constants $\alpha$ and $\beta_1$ play an important role in the amount of energy exchange ($\Delta \mathcal{E}$) between the fluid distributions in solution \ref{solA} while only constant $\beta_1$ has effect in solution \ref{solB}. 

    \begin{figure} 
        \centering
        \includegraphics[width=8cm,height=6.2cm]{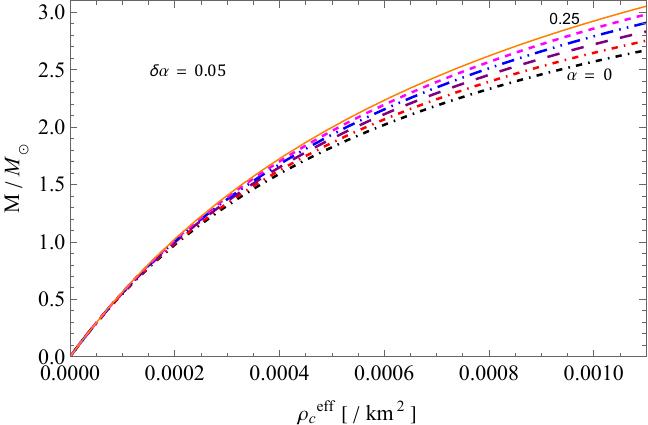}\\\includegraphics[width=8cm,height=6.2cm]{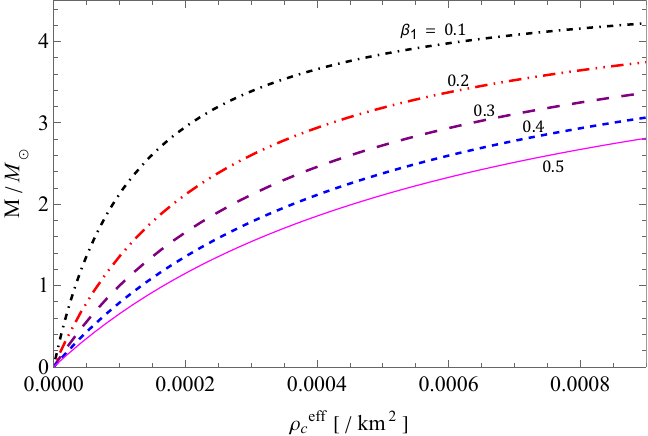}
        \caption{ The impact of constants $\alpha$-(left panel) and $\beta_1$ -(right panel) on mass-density ($m-\rho^{\text{eff}}$) curves that shows the stability to the solution \ref{solA} ($\rho=\theta^0_0$). } \label{m-rho-1b}
    \end{figure}
    \begin{figure} 
        \centering
        \includegraphics[width=8cm,height=6.2cm]{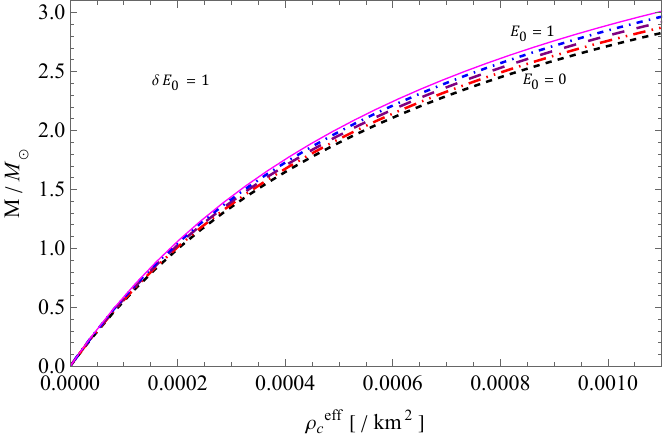}
        \caption{The impact of electric field on mass-density ($M-\rho_c^{\text{eff}}$) curves that shows the stability to the solution \ref{solA} ($\rho=\theta^0_0$).} \label{m-rho-1a-E}
    \end{figure}
    \begin{figure}
        \centering
        \includegraphics[width=8cm,height=6.2cm]{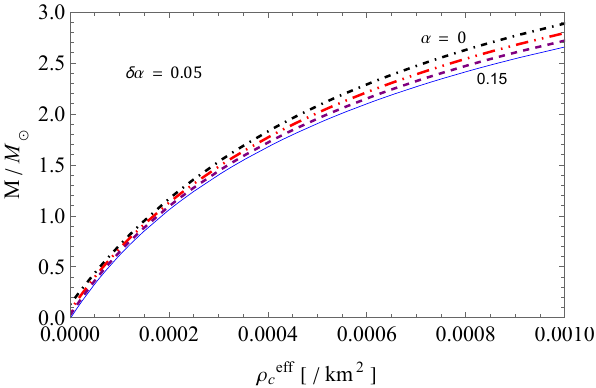}\\
        \includegraphics[width=8cm,height=6.2cm]{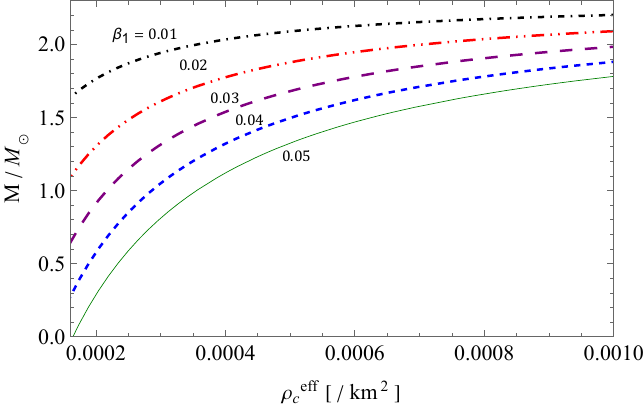}
        \caption{ The impact of constants $\alpha$-(left panel) and $\beta_1$ -(right panel) on mass-density ($M-\rho_c^{\text{eff}}$) curves that shows the stability to the solution \ref{solB} ($p_r=\theta^1_1$). } \label{m-rho-2a}
    \end{figure}
    \begin{figure}
        \centering
        \includegraphics[width=8cm,height=6.2cm]{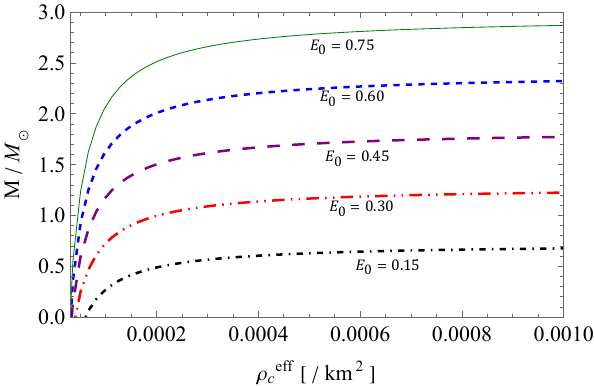}
        \caption{The impact of electric field on mass-density ($M-\rho_c^{\text{eff}}$) curves that shows the stability to the solution \ref{solB} ($p_r=\theta^1_1$).} \label{m-rho-2-E}
    \end{figure} 
    
    \begin{figure*}
        \centering
        \includegraphics[width=17cm,height=5.3cm]{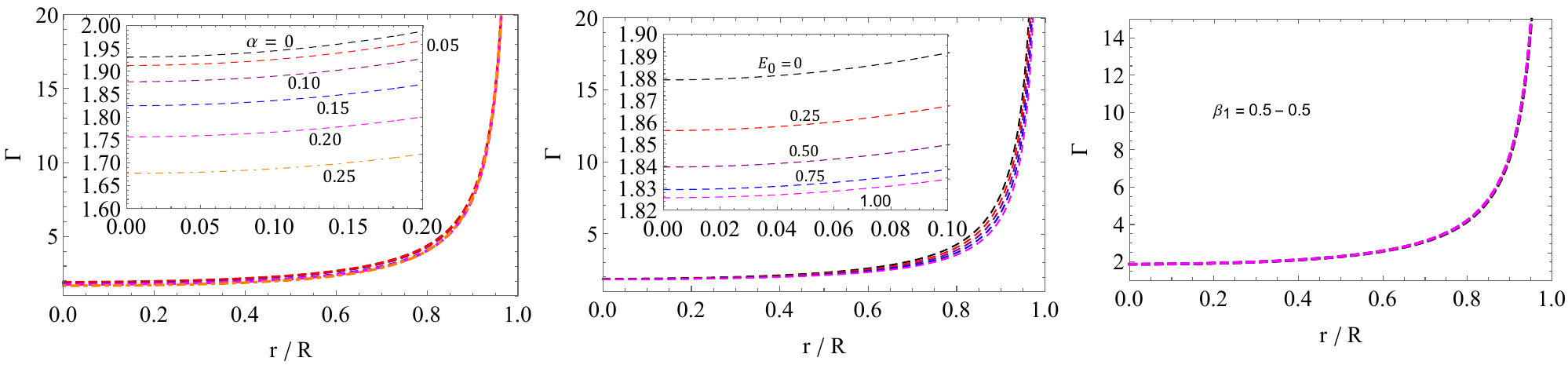}
        \caption{Variations of adiabatic index for solution \ref{solA} ($\rho=\theta^0_0$) with different values of $\alpha,~E_0$ and $\beta_1$.} \label{gam1}
    \end{figure*}

    \begin{figure*}
        \centering
        \includegraphics[width=17cm,height=5.5cm]{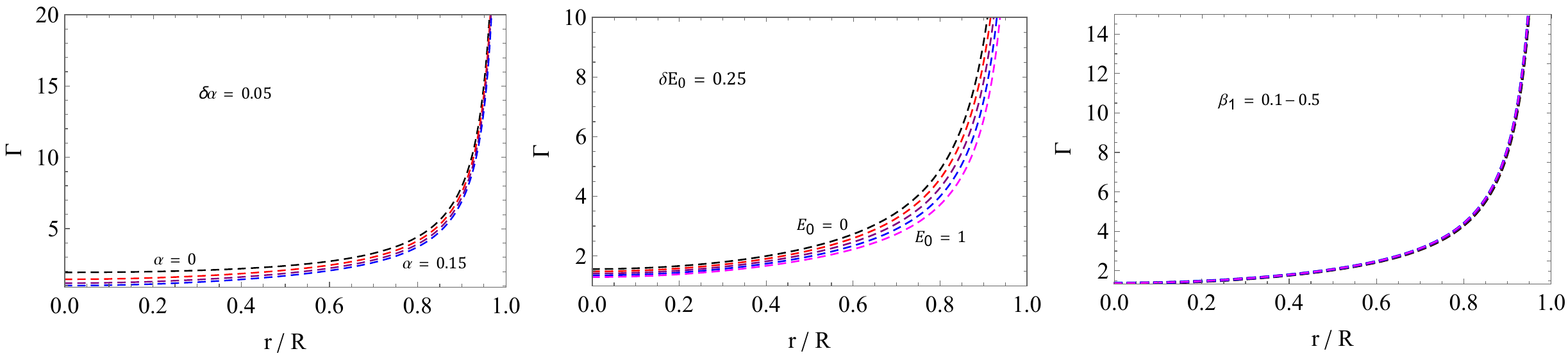}
        \caption{Variations of adiabatic index for solution \ref{solB} ($p_r=\theta^1_1$) with different values of $\alpha,~E_0$ and $\beta_1$.} \label{gam2}
    \end{figure*}

\section{Stability analysis} \label{sec8}
It is very important to check the stability of each configuration through small radial perturbation and Buchdahl's limit of gravitational collapse. This analysis can be performed via several techniques as below.

\subsection{Onset of gravitational collapse and modified Buchdahl limit}
In 1959, \citet{Buchdahl_1959_116} identified the onset of gravitational collapse of a spherically symmetric isotropic compact matter distribution. This configuration will trigger a gravitational collapse if $M/R$ is less than 4/9. However, this limit was extended for charged configurations by \citet{Andreasson_2009_288} as
    \begin{eqnarray}
        && \hspace{-4.7cm} {\frac{M}{R}} \leq {\frac{2R^2+3Q^2}{9R^2}}+{\frac{2}{9R}}\sqrt{R^2+3Q^2}. \label{eq86}
    \end{eqnarray}
This relationship can also be used to determine the critical value of surface charge just enough to avoid the collapse. equation \eqref{eq86} under small charge approximation i.e. $Q<R$ reduces to
    \begin{eqnarray}
        && \hspace{-5.4cm}{\frac{M}{R}} \lesssim {\frac{4}{9}}+{\frac{2Q^2}{3R^2}}+\mathcal{O}(Q^4/R^4). \label{eq87}
    \end{eqnarray}
Here one can see that the Buchdahl-Andr\'{e}asson limit is more than 4/9. Therefore, some net electric charge can inhibit gravitational collapse up to a certain limit. In a similar manner, a lower limit was also given for charge configurations by \cite{bonds} as
    \begin{eqnarray}
        && \hspace{-6.1cm} {\frac{Q^2(18R^2+Q^2)}{2R^2(12R^2+Q^2)}} \leq {\frac{M}{R}}.
    \end{eqnarray}
Generally, this lower bound has very small values, in our case its value is $\approx 10^{-4}$, hence the lower limit is satisfied. The satisfaction of Buchdahl-Andr\'{e}asson limit, i.e. the upper bound \eqref{eq87} also holds, which can be seen in Figs. \ref{m-r-1-a}, \ref{m-r-1-E}, \ref{m-r-2-a} and \ref{m-r-2-E}. Therefore, the presented solution is free from gravitational collapse.

\subsection{Stability under small radial perturbation}
Compact objects such as neutron stars and quark stars are stable for a lifetime of several million to several billion years. And therefore, any internal perturbation that arises from several exotic or non-exotic interactions must be stabilized by its own. This can be analyzed by adopting Chandrasekhar's method \citep{chand64} by perturbing the interior parameters like energy density, pressure, anisotropy etc. This treatment will be more handy if one considers a radial perturbation \textit{viz},
    \begin{eqnarray}
        && \hspace{-3.8cm} \Phi \rightarrow \Phi_0+\delta \Phi,~~ \nu \rightarrow \nu_0+\delta \nu, \nonumber \\
        && \hspace{-3.8cm} \rho^{\text{eff}} \rightarrow \rho^{\text{eff}}_0+\delta\rho^{\text{eff}},~~p^{\text{eff}}_r \rightarrow p^{\text{eff}}_{r0}+\delta p^{\text{eff}}_r \nonumber.
    \end{eqnarray}
Further, the perturbed components will be chosen as oscillatory function i.e. $\delta \mathcal{X} \sim e^{i\sigma t}$ ($\mathcal{X}\equiv \Phi,\, \nu, \, \rho^{\text{eff}},\, p^{\text{eff}}_r$), where $\sigma$ is the \textit{characteristic frequency}. As long as $\sigma^2 > 0$, the system is stable otherwise unstable. Harrison and his colleagues \citep{har65} and independently by \cite{zel71} adapted Chandrasekhar’s theorem into a much simpler technique. As per this new technique, it was found that $\sigma^2$ remains positive if the mass of the compact structure is an increasing function of its central density, i.e. $dM/d\rho_c^{\text{eff}}>0$. With the assumption of a polytropic matter i.e. $p^{\text{eff}}_r=\kappa (\rho^{\text{eff}})^\Gamma$ \big[$\Gamma$ is the adiabatic index\big], mass of the system is proportional to $[\rho_c^{\text{eff}}]^{3\{\Gamma-4/3\}/2}$, which can be an increasing function iff $\Gamma>4/3$. Therefore, for stable/non-collapsing compact structures, the mass should be an increasing function of the central density as well as the adiabatic index is $>4/3$. The configurations presenting in the current work satisfies this criteria, see Figs. \ref{m-rho-1b} and \ref{m-rho-1a-E} and therefore stable under radial perturbations. 

Moreover, the stability is enhanced by the strength of decoupling. In Fig. \ref{m-rho-1b} (left), for the same density range, the mass that can be supported by the system is more for $\alpha=0.25$ than $\alpha=0$. However, the opposite can be observed with the $f(Q)-$parameter $\beta_1$, i.e. less mass can be supported for large values of $\beta_1$ in the same density range (Fig. \ref{m-rho-1b} (right)). Next, the extra repulsive force due to the electric charge delays the gravitational instability that leads to a collapse and hence stability is gained in charged systems (Fig. \ref{m-rho-1a-E}).

\subsection{Onset of gravitational collapse}
In the above discussion on the stability of the compact stars under radial perturbations, it was required that the adiabatic index must be more than $4/3$. This is in fact true for isotropic fluid, and the same condition is different for anisotropic fluid. For the same Newtonian  fluid, the onset of gravitational collapse is inevitable when
    \begin{eqnarray}
        && \hspace{-5.8cm} \Gamma < {\frac{4}{3}} \left(1+{\frac{\Delta^{\text{eff}}}{r\, \left|\left(p^{\text{eff}}_{rc}\right)'\right|}}\right).
    \end{eqnarray}
Here it is clear that anisotropy increases the value of $\Gamma$ above 4/3, which depends on the strength of anisotropy. More the anisotropy easier is to trigger gravitational collapse. From the Figs. \ref{fig: anisotropy0} and \ref{fig: anisotropy1}, the anisotropy increases when $\alpha$ and $\beta_1$ increases. Therefore, as ($\alpha,~\beta_1$) increases the adiabatic index limit for collapse also shifted to higher values and therefore easier to initiate a collapse. At the same time increment in $\alpha$) reduces the value of $\Gamma$ at the core (See. Figs. \ref{gam1} (left) and \ref{gam2} (left)) and therefore approaching the $\Gamma_{\text{crit}}$ faster thus triggering collapse much more easier. Further, since $\beta_1$ has very minute effects on a core adiabatic index (See. Figs. \ref{gam1} (right) and \ref{gam2} (right)) and therefore the gravitational collapse will not trigger (or slower onset of collapse) until $\Gamma_{\text{crit}}>\Gamma_{\text{core}}$. Next, the effect of electric field strength parameter $\sqrt{E_0}$ have negligibly changes the $\Gamma_{\text{core}}$ for $\rho=\theta^0_0$ (See Fig. \ref{gam1} (mid)) however, for the second solution ($p_r=\theta^1_1$) $\Gamma_{\text{core}}$ decreases with increase in $\sqrt{E_0}$ (See Fig. \ref{gam2} (mid)). Therefore, increase in electric field strength may improve the stability under radial perturbations (See Fig. \ref{m-rho-2-E}), however, one must enhance it with the constraint $\Gamma_{\text{crit}}<\Gamma_{\text{core}}$, i.e. $\sqrt{E_0}$ has an upper limit.

\section{Discussion and conclusions} \label{sec9}
Using gravitational decoupling and the CGD technique, we now summarise our results and effort to simulate a strange star in $f(Q)$ gravity with an extra source under an electric field. We used the Tolman ansatz and the MIT bag model EOS for one of the gravitational potentials that served as a starting point to create a bounded star configuration. One family of solutions comes from the $\theta_0^0 = \rho$ sector, while the other comes from the $\theta_1^1 = p_r$ sector due to CGD formalism. In order to do this, we explored the impacts of anisotropy added using the complete Gravitational Decoupling approach and simulated compact objects within the context of the $f(Q)$ theory of gravity. By changing the decoupling parameters $\alpha$ and $\beta_1$ and the electric charge parameter $E_0$, these models were put through rigorous physical viability tests. Our assessments of our models' density and pressure profiles clearly show the effects of electric charge parameter $E_0$ and coupling parameter $\alpha$, $\beta_1$.\\

The motivation for the current study was to simulate compact objects within the context of $f(Q)$ gravity using gravitational wave occurrences (GW 190814) as well as observations of neutron stars PSR J1614-2230, PSR J1903+6620, Cen X-3 and LMC X-4. Our study explores the impact of anisotropy on compact stars by varying the decoupling constant based on widespread applications of the complete geometric deformation method. By employing the CGD method, we present the model of strange stars within the $f(Q)$ formalism in this investigation, and the mathematical model is validated through observational data from different compact stars, such as PSR J1614-2230, PSR J1903+6620, Cen X-3 and LMC X-4. A comparison has been made between our findings and the GW190814, an observational constraint on maximum allowable masses and radii for stable self-gravitating configurations.

For $f(Q)$ gravity, it is interesting that the presence of a secondary component of the GW 190814 event with mass $2.5-2.67\, \text{M}_{\odot}$ lying in the lower "\textit{mass gap}" in the absence of deformation has been achieved by using a decoupling constant $\alpha$ and slight deviation from parameter constant $\beta_1$. We have observed well-behaved mass profiles in both classes of solutions and demonstrated physically sound behaviour. Figs. \ref{m-r-1-a}, \ref{m-r-1-E}, \ref{m-r-2-a} and \ref{m-r-2-E} illustrate the novelty of our work. Although both solutions predict masses above $2\, \text{M}_{\odot}$, the observed neutron star limiting mass, they also predict masses above the masses of well-known compact objects like PSR J1614-2230, PSR J1903+6620, Cen X-3 and LMC X-4. For the $\theta_0^0 = \rho_0$ solution, the maximum masses span from $1.95\, \text{M}_{\odot}$ to $2.65\, \text{M}_{\odot}$ for $\alpha$ and from $1.52\, \text{M}_{\odot}$ to $3.35, \text{M}_{\odot}$ for $\beta_1$. The corresponding radii for these maximum masses range from 14.96 km to 15.64 km for $\alpha$ and from 10.71 km to 15.95 km for $\beta_1$. Regarding the $\theta_1^1=p_r$ solution, the radii for the maximum masses are between 14.64 km and 15.50 km for $\alpha$ and between 12.01 km and 18.05 km for $\beta_1$. The maximum masses in these models range from $2.3\, \text{M}_{\odot}$ to $2.7\, \text{M}_{\odot}$ for $\alpha$ and from $0.9\, \text{M}_{\odot}$ to $3.0\, \text{M}_{\odot}$ for $\beta_1$. We assessed the impact on stellar configuration regularity and stability by varying coupling parameters $\alpha$ and $\beta_1$. Our models displayed internal regularity, well-behaved energy density, and pressure profiles resembling realistic stars. The dominance of tangential pressure over radial pressure resulted in a repulsive anisotropic force stabilising the configuration. Stability analysis using the anisotropic Chandrasekhar adiabatic index and small perturbation criterion confirmed the stability (see Fig. \ref{m-rho-1b}, \ref{m-rho-1a-E}, \ref{m-rho-2a}, \ref{m-rho-2-E}), further enhanced by the decoupling constant. Increasing ($\alpha,~\beta_1$) raises anisotropy and facilitates collapse initiation, while higher $\alpha$ speeds up $\Gamma_{\text{crit}}$ approach. In Fig. \ref{gam1} and \ref{gam2}, we can see a negligible impact of $\beta_1$ on core collapse, but higher electric field ($\sqrt{E_0}$) may improve stability with $\Gamma_{\text{crit}}<\Gamma_{\text{core}}$ constraint. The satisfaction of Buchdahl--Andr\'easson's limit ensured the absence of gravitational collapse in our solution.

\section*{Data Availability}
There are no associated data with this article. No new data were generated or analysed in support of this research.

\section*{Acknowledgments}
SVL acknowledges the financial support provided by University Grants Commission (UGC) through Senior Research Fellowship  (UGC Ref. No.:191620116597) to carry out the research work. SKM appreciates the administration of the University of Nizwa in the Sultanate of Oman for their unwavering support and encouragement. SKM, KNS, BM acknowledge the support of IUCAA, Pune (India) through the visiting associateship program. AE thanks the National Research Foundation (NRF) of South Africa for the award of a postdoctoral fellowship. The authors are also thankful to Prof. Jorge Ovalle for his help in the coding of the contour diagrams.


\bibliographystyle{mnras}
\bibliography{references}
\bsp	
\label{lastpage}
\section{Appendix}
    \begin{small}
    \begin{eqnarray}
        && \hspace{-0.5cm} N_{11} (r) = \frac{1}{3 \beta_1 \left(C r^2+1\right) \left(C r^2+K\right)}\Big[2 \Big(-2 \mathcal{B}_g K \left(C r^2+1\right)^2-C^2 r^2 \nonumber\\
        && \hspace{0.5cm}\times \left(2 \beta_1+2 E_0 K-2 \beta_1 K+\beta_2 K r^2\right)+3 \beta_1 C (K-1)\nonumber\\
        && \hspace{0.5cm}-2 \beta_2 C K r^2-\beta_2 K\Big)\Big], \nonumber\\
        && \hspace{-0.5cm} N_{12} (r)=\frac{1}{3 \beta_1 \left(C r^2+1\right)^2 \left(C r^2+K\right)^2}\Big[2 \Big\{-2 \mathcal{B}_g K \left(C r^2+1\right)^2 \nonumber\\
        && \hspace{0.5cm} \times\left(3 C K r^2+C r^2 \left(C r^2-1\right)+K\right)+C^4 r^6 \big(2 E_0 K-2 \beta_1 (K-1)\nonumber\\
        && \hspace{0.5cm}+\beta_2 (-K) r^2\big)-C^3 r^4 \big(2 E_0 K (K+1)+\beta_1 \left(-2 K^2+9 K-7\right)\nonumber\\
        && \hspace{0.5cm}+\beta_2 K (3 K+1) r^2\big) +C^2 r^2 \big(-6 E_0 K^2+3 \beta_1 (K-1)^2+\beta_2 K \nonumber\\
        && \hspace{0.5cm}\times (1-7 K) r^2\big)+C K \left(3 \beta_1 (K-1)+\beta_2 (1-5 K) r^2\right)-\beta_2 K^2\Big\}\Big], \nonumber\\
        && \hspace{-0.5cm} N_{13}(r)=2 K \left(C r^2+1\right) \Big[4 (\alpha +1) \mathcal{B}_g K \left(C r^3+r\right)^2+C^2 r^4 \big(4 \alpha  \beta_1 \nonumber\\
        && \hspace{0.5cm}+\beta_1+4 (\alpha +1) E_0 K-4 (\alpha +1) \beta_1 K+2 (\alpha +1) \beta_2 K r^2\big)\nonumber\\
        && \hspace{0.5cm}+C r^2 \left(\beta_1 (6 \alpha -3 (2 \alpha +3) K+3)+4 (\alpha +1) \beta_2 K r^2\right)\nonumber\\
        && \hspace{0.5cm}+K \left(2 (\alpha +1) \beta_2 r^2-3 \beta_1\right)\Big],  \nonumber  \\
        && \hspace{-0.5cm}  N_{20}(r)=\alpha  \beta_1 \Big[\left(C r^2+1\right) \left(C r^2+K\right) \big(2 N_{12}(r)+N_{11}(r) \big((\alpha +2)\nonumber\\
        && \hspace{0.5cm} \times N_{11}(r) r^2+2\big)\big)-2 C N_{11}(r) (K-1) r^2\Big],\nonumber \\
        && \hspace{-0.5cm}   N_{21}(r)=2 K \left(C r^2+1\right) \Big[4 (\alpha +1) \mathcal{B}_g K \left(C r^3+r\right)^2+C^2 r^4 \big(4 \alpha  \beta_1+\beta_1 \nonumber\\
        && \hspace{0.5cm}+4 (\alpha +1) E_0 K-4 (\alpha +1) \beta_1 K+2 (\alpha +1) \beta_2 K r^2\big)\nonumber\\
        && \hspace{0.5cm}+C r^2 \left(\beta_1 (6 \alpha -3 (2 \alpha +3) K+3)+4 (\alpha +1) \beta_2 K r^2\right) \nonumber\\
        && \hspace{0.5cm}+K \left(2 (\alpha +1) \beta_2 r^2-3 \beta_1\right)\Big],\nonumber 
         \end{eqnarray}
    \begin{eqnarray}
        && \hspace{-0.5cm}  N_{22}(r)=\alpha  (\alpha +1) \beta_1 r^2 \left(C r^2+K\right) \Big[2 N_{12}(r)+N_{11}(r) \big((\alpha +1) \nonumber\\
        && \hspace{0.5cm} \times N_{11}(r) r^2+2\big)\Big] \Big[16 \mathcal{B}_g K \left(C r^2+1\right)^2+5 C^2 r^2 \big(2 \beta_1+2 E_0 K \nonumber\\
        && \hspace{0.5cm}-2 \beta_1 K+\beta_2 K r^2\big)-18 \beta_1 C (K-1)+10 \beta_2 C K r^2+5 \beta_2 K\Big], \nonumber 
    \end{eqnarray}
    \begin{eqnarray}
        && \hspace{-0.9cm} N_{33}(r)=  C \big[-6 (\alpha +1) \beta_1 \nonumber\\
        && \hspace{-0.1cm} \times \big(K \big(2 (\alpha -1) N_{12}(r) r^2+\left(\alpha ^2-1\right) N_{11}(r)^2 r^4+4 \alpha  N_{11}(r) r^2 \nonumber\\
        && \hspace{-0.1cm} +4\big) -2 (\alpha +1) N_{12}(r) r^2-\left((\alpha +1) N_{11}(r) r^2+2\right)^2\Big)+2 \beta_2 K r^2 \nonumber\\
        && \hspace{0.5cm} \times\big[2 \alpha ^2 r^2 \big\{N_{12}(r)+N_{11}(r) (N_{11}(r) r^2+2)\big\}+\alpha  \big(2 N_{12}(r) r^2\nonumber\\
        && \hspace{-0.1cm}+\left(N_{11}(r) r^2+2\right)^2\big)+\alpha ^3 N_{11}(r)^2 r^4+12\big]-3 \alpha  E_0 K \big\{2 (\alpha +1)\nonumber\\
        && \hspace{-0.1cm} \times N_{12}(r) r^2+\big((\alpha +1) N_{11}(r) r^2+2\big)^2\big\}\big]+K \Big\{4 (\alpha +3) \beta_2\nonumber\\
        && \hspace{0.5cm}+2 (\alpha +1) N_{12}(r) \left(6 \beta_1+\alpha  \beta_2 r^2\right)+(\alpha +1)^2 N_{11}(r)^2 r^2 \nonumber\\
        && \hspace{0.5cm} \times \left(6 \beta_1+\alpha  \beta_2 r^2\right)+4 (\alpha +1)  N_{11}(r) \left(3 \beta_1+\alpha  \beta_2 r^2\right)\Big\}
    \end{eqnarray}
    \end{small}

\end{document}